\documentclass[twocolumn]{aastex631}

\newcommand{\avir}{\mbox{$\alpha_{\rm vir}$}}
\newcommand{\kms}{\mbox{km~s$^{-1}$}}
\newcommand{\twco}{\mbox{$^{12}$CO}}
\newcommand{\ttco}{\mbox{$^{13}$CO}}

\newcommand{\HII}{\mbox{H\thinspace {\sc ii}}}
\newcommand{\CI}{\mbox{C\thinspace {\sc i}}}
\newcommand{\CII}{\mbox{C\thinspace {\sc ii}}}
\def\farcs{\hbox{$.\!\!^{\prime\prime}$}}
\defcitealias{solomon:87}{S87}

\received{2022 January 23}
\accepted{2022 May 23}
\published{2022 June 15}

\shorttitle{30 Doradus Molecular Cloud}
\shortauthors{Wong et al.}
\graphicspath{{./}{figures/}}

\begin{document}

\title{The 30 Doradus Molecular Cloud at 0.4 pc Resolution with the Atacama Large Millimeter/submillimeter Array: Physical Properties and the Boundedness of CO-emitting Structures}

\correspondingauthor{Tony Wong}
\email{wongt@illinois.edu}

\author[0000-0002-0786-7307]{Tony Wong}
\affiliation{Astronomy Department, University of Illinois, Urbana, IL 61801, USA}

\author{Luuk Oudshoorn}
\affiliation{Leiden Observatory, Leiden University, Niels Bohrweg 2, 2333CA Leiden, The Netherlands}

\author{Eliyahu Sofovich}
\affiliation{Astronomy Department, University of Illinois, Urbana, IL 61801, USA}

\author[0000-0002-8432-3362]{Alex Green}
\affiliation{Astronomy Department, University of Illinois, Urbana, IL 61801, USA}

\author{Charmi Shah}
\affiliation{Astronomy Department, University of Illinois, Urbana, IL 61801, USA}

\author[0000-0002-4663-6827]{R\'emy Indebetouw}
\affiliation{Department of Astronomy, University of Virginia, P.O. Box 3818, Charlottesville, VA 22903, USA}
\affiliation{National Radio Astronomy Observatory, 520 Edgemont Rd., Charlottesville, VA, 22903, USA}

\author[0000-0002-0522-3743]{Margaret Meixner}
\affiliation{SOFIA-USRA, NASA Ames Research Center, MS 232-12, Moffett Field, CA 94035, USA}

\author[0000-0001-5397-6961]{Alvaro Hacar}
\affiliation{Department of Astrophysics, University of Vienna, Türkenschanzstrasse 17, 1180 Vienna, Austria}

\author[0000-0001-6576-6339]{Omnarayani Nayak}
\affiliation{Space Telescope Science Institute, 3700 San Martin Drive, Baltimore, MD 21218, USA}

\author[0000-0002-2062-1600]{Kazuki Tokuda}
\affil{Department of Earth and Planetary Sciences, Faculty of Sciences, Kyushu University, Nishi-ku, Fukuoka 819-0395, Japan}
\affil{National Astronomical Observatory of Japan, National Institutes of Natural Sciences, 2-21-1 Osawa, Mitaka, Tokyo 181-8588, Japan}
\affil{Department of Physics, Graduate School of Science, Osaka Metropolitan University, 1-1 Gakuen-cho, Naka-ku, Sakai, Osaka 599-8531, Japan}

\author[0000-0002-5480-5686]{Alberto D. Bolatto}
\affiliation{Department of Astronomy and Joint Space Science Institute, University of Maryland, College Park, MD 20742, USA}
\affiliation{Visiting Astronomer, National Radio Astronomy Observatory, Charlottesville, VA 22903, USA}

\author[0000-0002-5635-5180]{M\'elanie Chevance}
\affiliation{Astronomisches Rechen-Institut, Zentrum f\"{u}r Astronomie der Universit\"{a}t Heidelberg, M\"{o}nchhofstra\ss e 12-14, D-69120 Heidelberg, Germany}

\author[0000-0001-7906-3829]{Guido De Marchi}
\affiliation{European Space Research and Technology Centre, Keplerlaan 1, 2200 AG Noordwijk, Netherlands}

\author[0000-0002-8966-9856]{Yasuo Fukui}
\affil{Department of Physics, Nagoya University, Chikusa-ku, Nagoya 464-8602, Japan}

\author[0000-0002-2954-8622]{Alec S.\ Hirschauer}
\affiliation{Space Telescope Science Institute, 3700 San Martin Drive, Baltimore, MD 21218, USA}

\author[0000-0001-7105-0994]{K.~E.~Jameson}
\affiliation{CSIRO, Space and Astronomy, PO Box 1130, Bentley, WA 6102, Australia}

\author{Venu Kalari}
\affiliation{Gemini Observatory, NSF NOIRLab, Casilla 603, La Serena, Chile}

\author[0000-0002-7716-6223]{Vianney Lebouteiller}
\affiliation{AIM, CEA, CNRS, Université Paris-Saclay, Université Paris Diderot, Sorbonne Paris Cité, 91191 Gif-sur-Yvette, France}

\author[0000-0002-4540-6587]{Leslie W. Looney}
\affiliation{Astronomy Department, University of Illinois, Urbana, IL 61801, USA}

\author[0000-0003-3229-2899]{Suzanne C. Madden}
\affiliation{Departement d'Astrophysique AIM/CEA Saclay, Orme des Merisiers, 91191 Gif-sur-Yvette, France}

\author[0000-0001-7826-3837]{Toshikazu Onishi}
\affil{Department of Physics, Graduate School of Science, Osaka Metropolitan University, 1-1 Gakuen-cho, Naka-ku, Sakai, Osaka 599-8531, Japan}

\author[0000-0001-6326-7069]{Julia Roman-Duval}
\affiliation{Space Telescope Science Institute, 3700 San Martin Drive, Baltimore, MD 21218, USA}

\author[0000-0002-5307-5941]{M\'onica Rubio}
\affiliation{Departamento de Astronom\'ia, Universidad de Chile, Casilla 36-D, Santiago, Chile}

\author[0000-0003-0306-0028]{A.G.G.M.~Tielens}
\affiliation{Department of Astronomy, University of Maryland, College Park, MD 20742, USA}
\affiliation{Leiden Observatory, Leiden University, Niels Bohrweg 2, 2333CA Leiden, The Netherlands}

\begin{abstract}

We present results of a wide-field (approximately 60 $\times$ 90 pc) ALMA mosaic of CO(2--1) and $^{13}$CO(2--1) emission from the molecular cloud associated with the 30 Doradus star-forming region.  
Three main emission complexes, including two forming a bowtie-shaped structure extending northeast and southwest from the central R136 cluster, are resolved into complex filamentary networks. 
Consistent with previous studies, we find that the central region of the cloud has higher line widths at fixed size relative to the rest of the molecular cloud and to other LMC clouds, indicating an enhanced level of turbulent motions.  
However, there is no clear trend in gravitational boundedness (as measured by the virial parameter) with distance from R136.  
Structures observed in \ttco\ are spatially coincident with filaments and are close to a state of virial equilibrium.  
In contrast, \twco\ structures vary greatly in virialization, with low CO surface brightness structures outside of the main filamentary network being predominantly unbound.  
The low surface brightness structures constitute $\sim$10\% of the measured CO luminosity; they may be shredded remnants of previously star-forming gas clumps, or alternatively the CO-emitting parts of more massive, CO-dark structures.

\end{abstract}

\keywords{galaxies: ISM --- radio lines: ISM --- ISM: molecules --- Magellanic Clouds}

\section{Introduction} \label{sec:intro}

As the most luminous star forming region in the Local Group, the supergiant \HII\ region of the Large Magellanic Cloud known as the Tarantula Nebula or 30 Doradus (hereafter 30 Dor) provides a unique opportunity to study massive star formation and how it drives and responds to stellar feedback.
At the heart of 30 Dor lies R136, a young ($\sim$1--2 Myr; \citealt{crowther:16,bestenlehner:20}) compact ($r\sim1$ pc) star cluster with extraordinarily high stellar densities of $> 1.5\times 10^4$ M$_\odot$ pc$^{-3}$ \citep{selman:13} and containing several stars with initial masses exceeding the canonical stellar mass upper limit of 150 M$_\odot$ \citep{crowther:10}.
\citet{bestenlehner:20} find that R136 alone contributes $\sim$27\% of the ionizing flux and $\sim$19\% of the overall mechanical feedback in 30 Dor (as measured within a 150 pc radius by \citealt{doran:13}).
On larger scales, the cumulative impact of stellar winds and supernova explosions is apparent in the $\sim$3--9 $\times 10^6$ K plasma responsible for diffuse X-ray emission \citep{townsley:06}. 
The rich observational data for 30 Dor have been complemented by extensive theoretical modeling of the associated \HII\ and photon dominated regions \citep[e.g.,][]{lopez:11,pellegrini:11,chevance:16,chevance:20,rahner:18}.
As a result, 30 Dor is a promising local analogue for the extreme conditions that were common during the peak epoch of star formation in the Universe.

R136 and its immediate surroundings have traditionally received the most attention, however it has become clear that star formation is on-going in the giant molecular cloud beyond the central cluster \citep[e.g.,][]{walborn:13}.
A spatially extended distribution of upper main sequence stars was found by the {\it Hubble} Tarantula Treasury Program (HTTP) survey, which imaged a $14^\prime \times 12^\prime$ (200 $\times$ 175 pc) region of 30 Dor to characterize the stellar populations and to derive a dust extinction map using stellar photometry \citep{sabbi:13,sabbi:16,demarchi:16}.
The distribution and ages of O and B stars, as determined by the VLT-FLAMES Tarantula Survey, also indicate that massive star formation has been widely distributed throughout 30 Dor \citep{schneider:18}.
The discovery of $\sim$20\,000 pre-main sequence (PMS) stars using HTTP photometry \citep{ksoll:18}, together with the $\sim$40 embedded massive young stellar objects (YSOs) previously discovered by the {\it Spitzer} SAGE \citep{whitney:08,gruendl:09} and {\it Herschel} HERITAGE \citep{seale:14} programs, have made 30 Dor one of the best studied regions of current star formation activity in any galaxy.

In contrast to the stellar population and PMS/YSO studies, available molecular gas maps of the 30 Dor region have much poorer angular resolution \citep[$\gtrsim$10 pc;][]{johansson:98,minamidani:08,wong:11,kalari:18,okada:19}, aside from previously published data from the Atacama Large Millimeter/submillimeter Array (ALMA) covering a relatively small (12 $\times$ 12 pc) area \citep{indebetouw:13,indebetouw:20}.
To address these limitations, we have conducted new observations with ALMA, exploiting the array's unique capability to obtain a sensitive, high-resolution (1\farcs75 beam) map of the giant molecular cloud complex across an extent of $\sim$100 pc using the CO $J$=2--1 and $^{13}$CO $J$=2--1 transitions.
These low-$J$ CO transitions can be used to probe the molecular gas column density and turbulent properties down to sub-parsec scales at a spectral resolution of $\sim$0.1 km s$^{-1}$, with the important caveat that the ability of CO to trace H$_2$ may be affected by the low metallicity and strong radiation field in this region \citep{israel:97,bolatto:13a,jameson:16,chevance:20}.

In this paper we present the basic ALMA data products (\S\ref{sec:obs}, \S\ref{sec:mom}) and characterize the CO and \ttco\ emission structures using dendrogram (\S\ref{sec:dendro}) and filament finding (\S\ref{sec:filfinder}) approaches.
Our immediate goal is to revisit, over a much larger region, results from previous ALMA studies \citep{indebetouw:13,nayak:16,wong:17,wong:19} which have found that the CO line width is enhanced in the 30 Dor region relative to molecular clouds in the Milky Way or elsewhere in the LMC.
In \S\ref{sec:results} we examine whether this enhancement is found throughout the 30 Dor region and how it relates to the gravitational boundedness of molecular gas structures.
We briefly summarize and discuss our results in \S\ref{sec:disc}.
In related works, we will present a greatly expanded catalog of YSOs across the ALMA field and examine the relationship between CO emission and YSOs (O. Nayak et al., submitted), and we will conduct a comparative study to examine the effect of local star formation activity (as probed by mid-infrared brightness) on molecular cloud properties across the LMC (A. Green et al., in preparation).  We adopt an LMC distance of 50 kpc \citep{pietrzynski:19} throughout this paper, for which 1\arcmin\ is equivalent to 14.5 pc and 1\arcsec\ is equivalent to 0.24 pc.

\section{Observations and Data Reduction}\label{sec:obs}

The data presented in this paper were collected for ALMA Cycle 7 project 2019.1.00843.S in 2019 October to December.  Since the field is larger than can be observed in a single ALMA scheduling block, it was split into five rectangular subfields that were observed and imaged separately.  To recover flux across the widest possible range of spatial scales, each subfield was observed in the ALMA ACA (hereafter 7m) and Total Power (hereafter TP) arrays in addition to the compact (C43-1) configuration of the 12m array.  Four of the subfields spanned 150\arcsec\ $\times$ 150\arcsec\ and consisted of 149 individual pointings of the 12m array, observed for about 20 sec per pointing, and 52 pointings of the 7m array, observed for about 7 min per pointing.   The fifth subfield in the northeast was half the size of the others (150\arcsec\ $\times$ 75\arcsec).  Nearly all data used J0601-7036 as the phase calibrator, which varied between 220 and 300 mJy during the span of observations.  Absolute flux calibration was set using the observatory-monitored quasar grid, specifically one of the sources J0519-4546, J0538-4405, or J1107-5509 for each execution of the project. The correlator was set to cover the CO ($J$=2--1) and $^{13}$CO ($J$=2--1) lines at high ($\sim$0.1 \kms) spectral resolution, the C$^{18}$O ($J$=2--1) and H$_2$CO ($3_{2,1}$-$2_{2,0}$, $3_{2,2}$-$2_{2,1}$, and $3_{0,3}$-$2_{0,2}$) lines at moderate ($\sim$0.4 \kms) spectral resolution, and the H30$\alpha$ and continuum across a 1.9 GHz window at low ($\sim$1.5 \kms) spectral resolution.  For the 12m data the time-varying gains were transferred from the wide to narrow spectral windows, and for the 7m data, all spectral windows were combined to solve for time-varying gain. In this paper we focus on the results of the CO and $^{13}$CO observations; a study of the H$_2$CO emission will appear separately (Indebetouw et al., in preparation).

\begin{figure*}
\begin{center}
\includegraphics[width=0.6\textwidth]{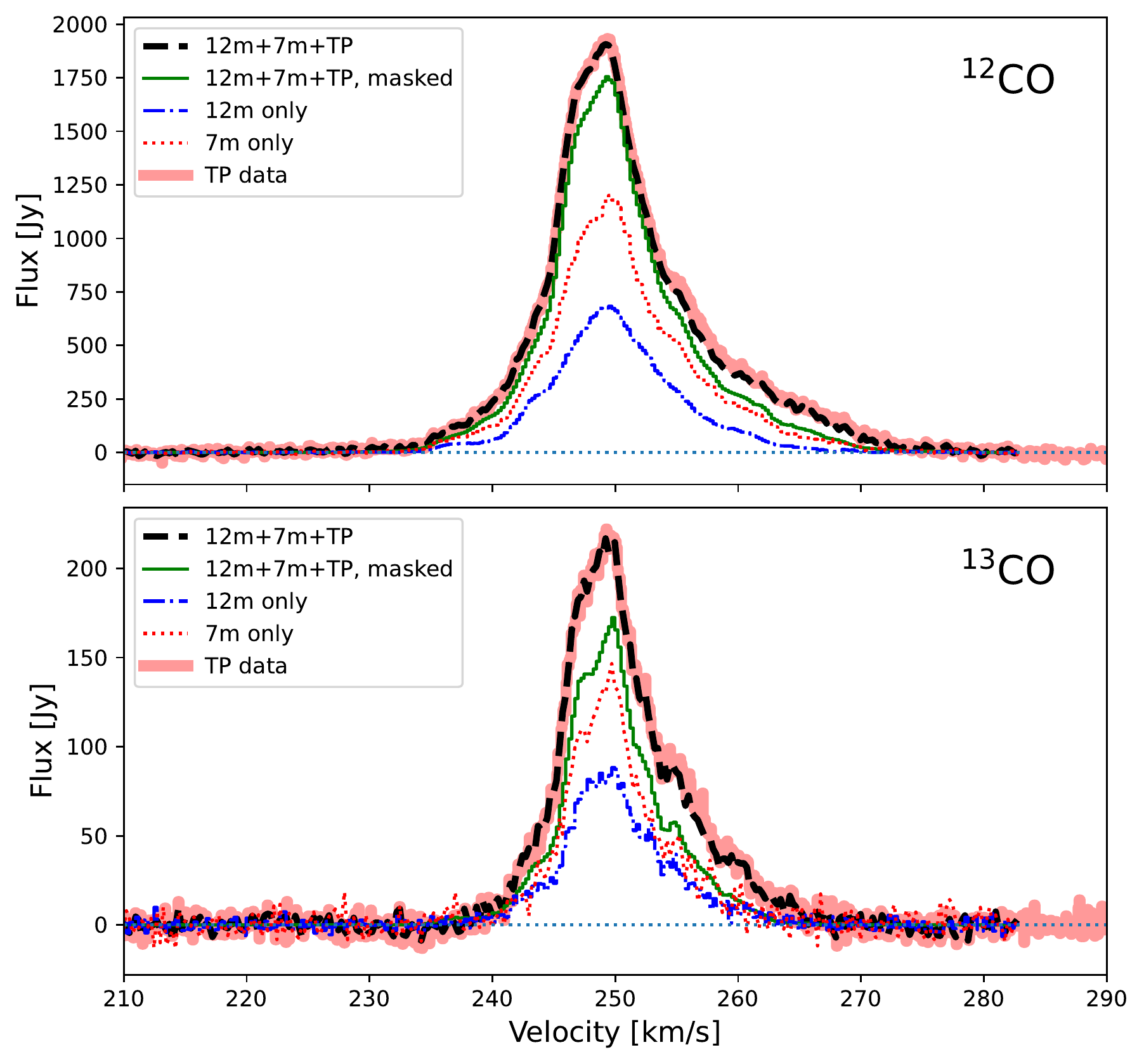}
\end{center}
\caption{Integrated flux spectra for the CO(2--1) (top) and $^{13}$CO(2--1) (bottom) cubes at 0.25 \kms\ resolution.  The cubes compared are the feathered cube ({\it black dashed line}), {the TP array data only ({\it thick pink line}),} the 7m array data only ({\it red dotted line}), and the 12m array data only ({\it blue dot-dashed line}).  A solid green line shows the flux in the feathered cubes after applying the dilated mask described in \S\ref{sec:mom}.
\label{fig:fluxcomp}}
\end{figure*}

Visibilities were calibrated by the observatory staff using Pipeline-CASA56-P1-B and CASA 5.6.1-8, with 
imaging then performed in CASA 5.6.1.
For the TP data, the {\tt sdimaging} task was used to generate image cubes from the spectra.  A residual sinusoidal baseline in the $^{13}$CO TP cube was removed from the gridded image cube: at each position, the line-free frequency ranges of a spectrum averaged over a 60$\arcsec$ square region were fitted with two sinusoids of different period and amplitude, and the resulting  baseline subtracted.  The dominant effect on the image cube is to remove modest off-source negative bowls. For the 7m and 12m data, the {\tt uvcontsub} task was first used to subtract the continuum using a 0-order fit to line-free channels (conservatively chosen based on previous imaging).  The {\tt tclean} task was then used to generate image cubes with a Briggs robustness parameter of 0.5, a threshold of 0.18 mJy, and a restoring beam of 1\farcs75 FWHM for the 12m data (7\arcsec\ FWHM for the 7m data).  After cleaning, the 7m and TP cubes were combined using the {\tt feather} task, and the 12m and 7m+TP cubes were combined using a second run of {\tt feather}.  Since the sensitivity pattern for each subfield has a decreasing extent in going from TP to 7m to 12m, each feathering step was performed on images tapered by the narrower sensitivity pattern (7m in the first step, 12m in the second) and the final results are assumed to have the sensitivity pattern of the 12m images.  

Figure~\ref{fig:fluxcomp} compares the integrated spectra derived from the 12m and 7m data alone with those derived from the TP data and from the feathering process.  The velocity axis uses the radio definition of velocity, $c(\nu_0-\nu)/\nu_0$, and is referenced to the kinematic Local Standard of Rest (LSR).  As expected, the TP flux (shown as the thick pink line) is recovered in the feathered cube (shown as the dashed black line).  Flux recovery for the 7-meter (12-meter) array alone is 60\% (33\%) for \twco\ and 55\% (38\%) for \ttco.  The threshold mask used to construct the moment images (shown as the green line; see \S\ref{sec:mom}) recovers $\sim$80\% of the feathered \twco\ flux and $\sim$70\% of the feathered \ttco\ flux; the remaining flux lies outside the mask boundary.  The integrated \twco\ TP flux is 22900 Jy \kms, which corresponds to a molecular gas mass (including helium) of $2.4 \times 10^5\;M_\odot$ for our adopted distance and CO-to-H$_2$ conversion factor (\S\ref{sec:mom}). 

To generate the final maps, gain-corrected image cubes for each subfield were mosaiced by co-addition using inverse variance weighting based on the sensitivity pattern of each subfield.  The mosaicing was performed using the Python {\sc reproject} package\footnote{\url{https://reproject.readthedocs.io/}} using bilinear interpolation.  After mosaicing, the images were downsampled by a factor of two in RA and DEC to yield final images of 1000 $\times$ 800 pixels using 0\farcs5 pixels; this is still more than adequate to oversample the 1\farcs75 synthesized beam (corresponding to 0.4 pc at our adopted distance).  In addition to cubes with 0.1 \kms\ channels (spanning 200 to 289.9 \kms), we also generated cubes with 0.25 \kms\ channels (spanning 208 to 282 \kms) to improve the brightness sensitivity per channel.  The resulting rms noise per 0.25 \kms\ channel is $\approx$0.26 K (35 mJy beam$^{-1}$), with somewhat lower noise ($\approx$0.16 K or 21 mJy beam$^{-1}$) in the smallest subfield.  Most of the results in this paper are based on analysis of the 0.25 \kms\ cubes, though comparisons with the 0.1 \kms\ cubes are made as well.

\section{Data Analysis Methods}

\begin{figure*}
\includegraphics[width=\textwidth]{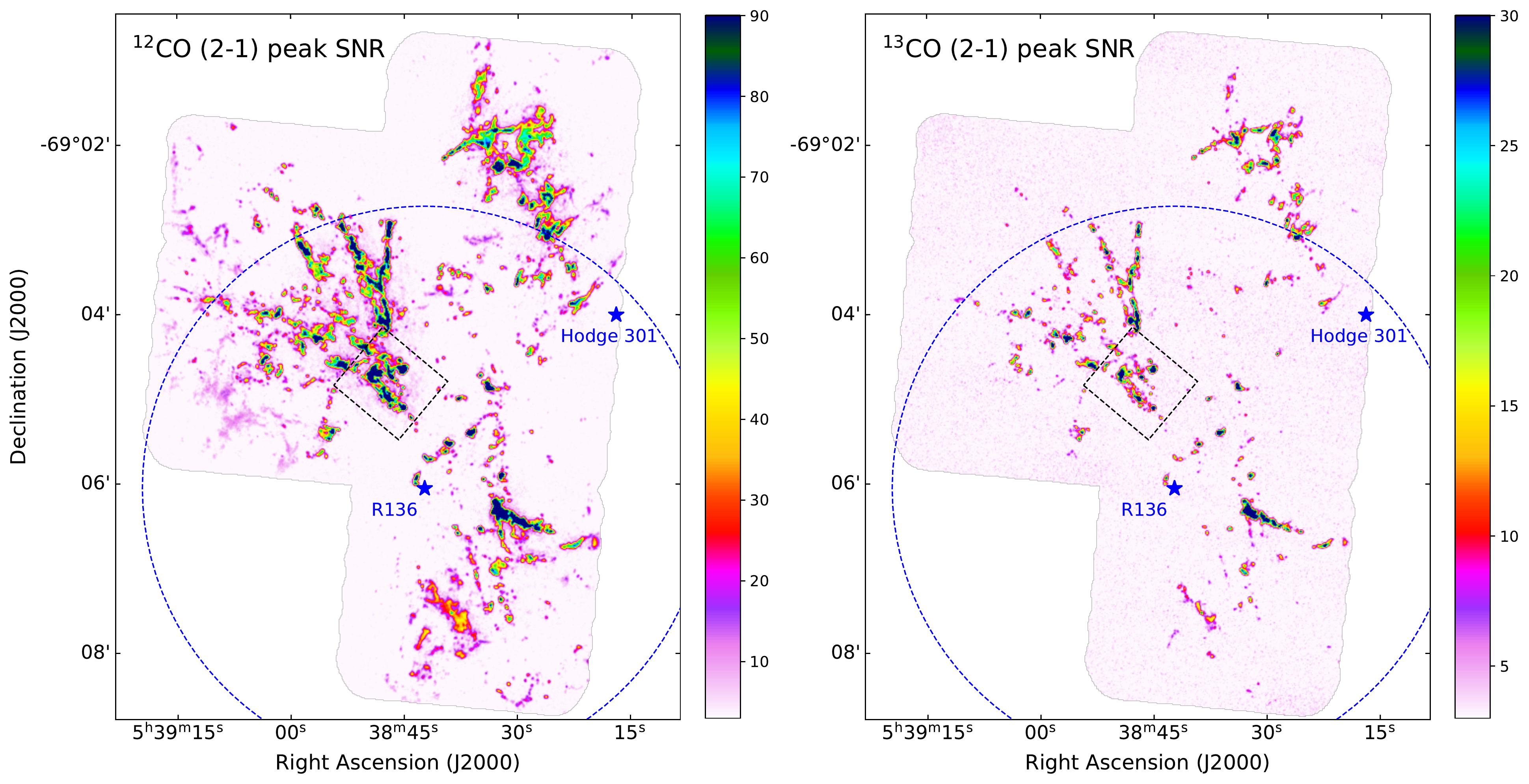}
\caption{Peak SNR images for the CO (left) and $^{13}$CO (right) cubes. The dashed circle represents a projected distance of 200\arcsec\ (48 pc) from the center of the R136 cluster, for ease of comparison with Fig.~\ref{fig:refdist_alpha}.  The dashed rectangle has a linear dimension of $\sim$12 pc and denotes the region mapped in ALMA Cycle 0 \citep{indebetouw:13}.  The central position of the more evolved Hodge 301 cluster is also indicated.
\label{fig:snrpk}}
\end{figure*}

\begin{figure*}
\begin{center}
\includegraphics[angle=90,width=3.25in]{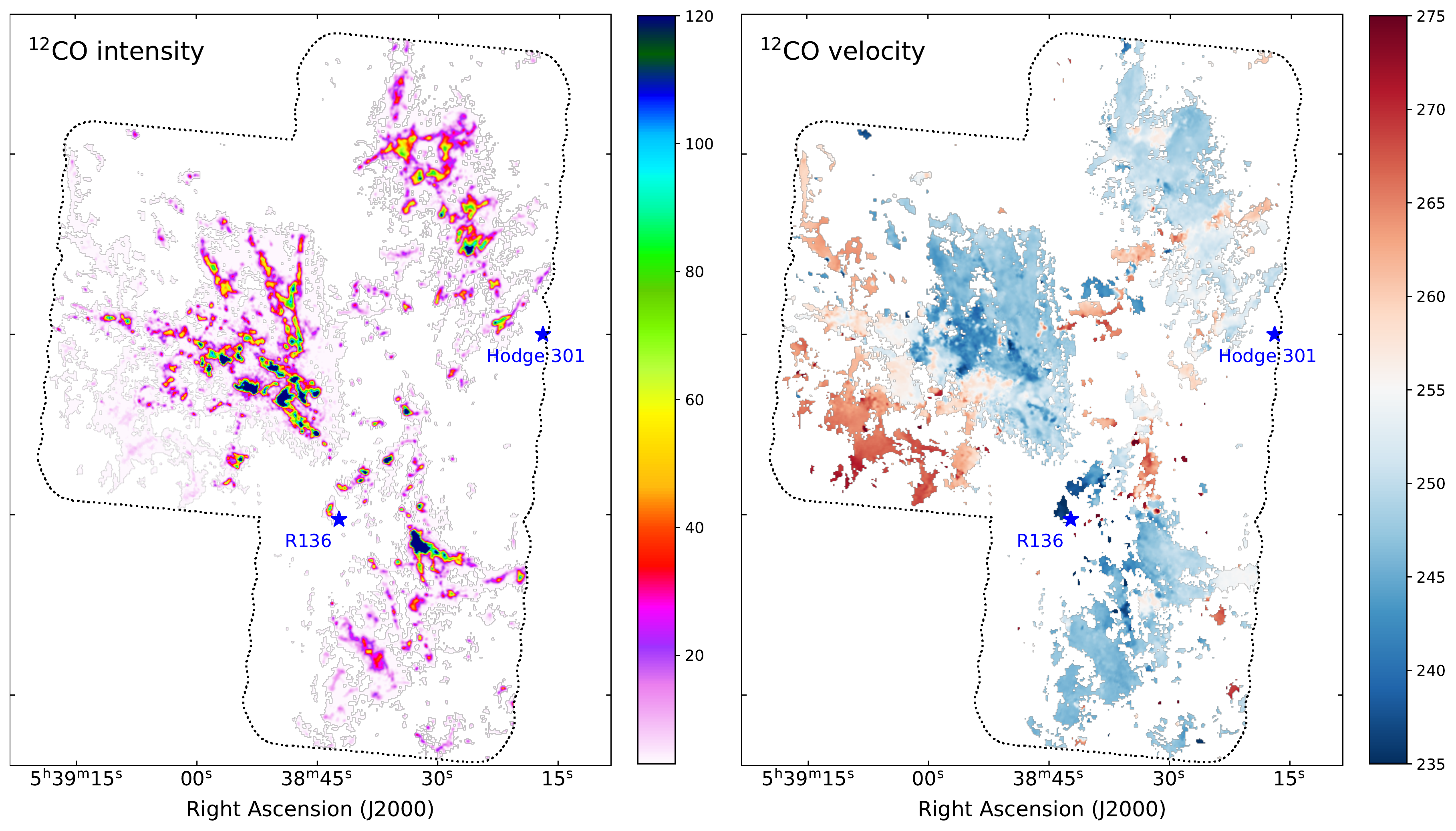}\\[1ex]
\includegraphics[angle=90,width=3.25in]{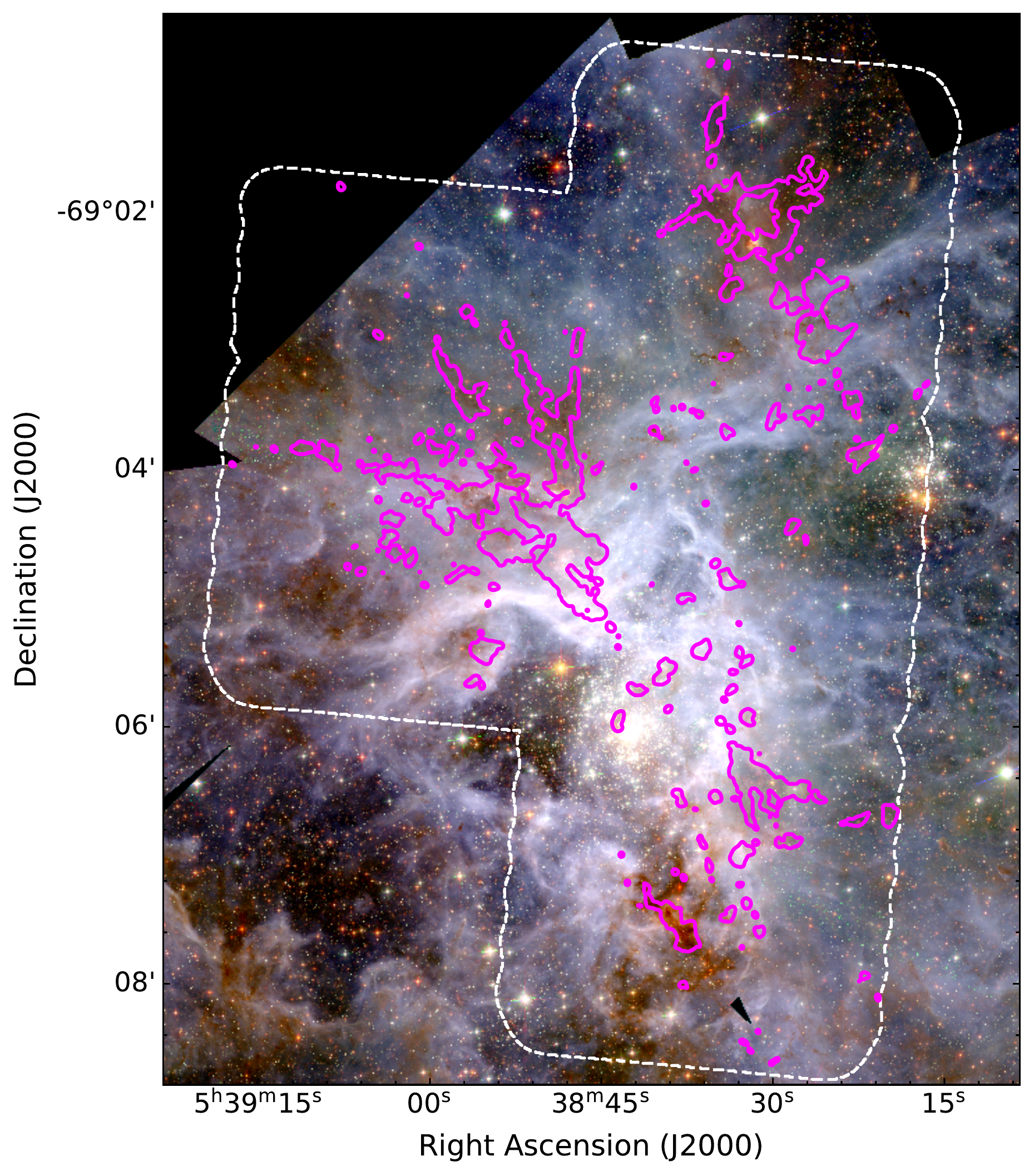}
\end{center}
\caption{0th moment (integrated intensity in K \kms, {middle}) and 1st moment (intensity-weighted mean velocity in \kms, right) images for the CO cube, after applying the dilated mask. The outline of the ALMA footprint is indicated by a dotted contour.  {In the left panel, the 0th moment contours are overlaid on a {\it Hubble Space Telescope} RGB image from the HTTP survey \citep{sabbi:13} with 1.6 $\mu$m in red, 775 nm in green, and 555 nm in blue.}
\label{fig:mom01}}
\end{figure*}

\subsection{Intensities and Column Densities}\label{sec:mom}

Figure~\ref{fig:snrpk} shows images of peak signal-to-noise ratio (SNR) for the \twco\ and \ttco\ data with 0.25 \kms\ channels.  Although insensitive to complex line profiles, such images effectively reveal the full dynamic range of detected emission without requiring subjective decisions about how to mask out noise.  For this reason the peak SNR image for \twco\ is used for filament identification in \S\ref{sec:filfinder}.  The dashed circle is at a projected distance of $\theta_{\rm off}$=200\arcsec\ from the center of the R136 cluster at 
$\alpha_{2000}$=5$^{\mathrm h}$38$^{\mathrm m}$42\fs3,
$\delta_{2000}$=$-69$\arcdeg06\arcmin03\farcs3 \citep{sabbi:16}.
The central position of the older Hodge 301 cluster ($\alpha_{2000}$=5$^{\mathrm h}$38$^{\mathrm m}$17$^{\mathrm s}$, $\delta_{2000}$=$-69$\arcdeg04\arcmin00\arcsec; \citealt{sabbi:16}) is indicated as well.

We have also generated intensity moment images from the cubes, using a signal masking procedure implemented in the Python {\tt maskmoment} package.\footnote{\url{https://github.com/tonywong94/maskmoment}}  In brief, starting from a gain-corrected cube and an rms noise cube, a strict mask composed of pixels with brightness of $4\sigma$ or greater in two consecutive channels is created and expanded to a looser mask defined by the surrounding $2\sigma$ contour.  Mask regions with projected sky area less than two synthesized beams are then eliminated.  The resulting integrated flux spectrum within the mask is shown as the green line in Figure~\ref{fig:fluxcomp}.  The 0th, 1st, and 2nd intensity moments along the velocity axis are then computed with pixels outside the signal mask blanked.  Images of the 0th and 1st moments of the \twco\ cube are shown in Figure~\ref{fig:mom01}.  A notable feature of the 1st moment map is the roughly orthogonal blueshifted and redshifted emission structures that are found crossing the center of the map.  We provide an overview of the CO distribution and velocity structure in \S\ref{sec:overview}.

Derivation of molecular gas mass from the cubes follows the basic procedures presented in \citet{wong:17} and \citet{wong:19}.  Where \ttco\ emission is detected, we can determine the \ttco\ column density in the local thermodynamic equilibrium (LTE) approximation, $N(\ttco)$.  The excitation temperature $T_{\rm ex}$ is assumed constant along each line of sight and is derived from the \twco\ peak brightness temperature ($T_{\rm 12, pk}$) by assuming the \twco\ line is optically thick at the peak of the spectrum and is not subject to beam dilution:
\begin{equation}\label{eqn:t12}
T_{\rm 12, pk} = J(T_{\rm ex}) - J(T_{\rm cmb})\;,
\end{equation}
where
\begin{equation}
J(T) \equiv \frac{h\nu/k}{\exp(h\nu/kT)-1}\;.
\end{equation}
For pixels with \ttco\ peak SNR $>$5, the median and maximum values of $T_{\rm ex}$ are found to be 20 K and 60 K respectively.  The beam-averaged \ttco\ optical depth, $\tau_{13}$, is then calculated from the brightness temperature, $T_{13}$, at each position and velocity in the cube by solving
\begin{equation}\label{eqn:t13}
T_{13} = [J(T_{\rm ex}) - J(T_{\rm cmb})][1-\exp(-\tau_{13})]\;.
\end{equation}
As noted in \citet{wong:17} and \citet{wong:19}, $T_{13}$ cannot exceed $J(T_{\rm ex}) - J(T_{\rm cmb}) \approx T_{\rm ex}-4.5$ (approximation good to 0.8 K for $5<T_{\rm ex}<60$).
Adopting a minimum value for the excitation temperature serves to reduce the number of undefined values of $\tau_{13}$ and prevents noise in the \ttco\ map from being assigned very large opacities. We adopt a minimum $T_{\rm ex} = 8$ K under the assumption that lower inferred values of $T_{\rm ex}$ result from beam dilution of \twco.  Since only 1.1\% of highly significant (\ttco\ peak SNR $>$ 5) pixels fall below this limit, our results are not sensitive to this choice.
The inferred column density $N(\ttco)$ in cm$^{-2}$, summed over all rotational levels, is determined from $T_{\rm ex}$ and $\tau_{13}$ using the equation \citep[e.g.,][Appendix A]{Garden:91}:
\begin{equation}
    N(\ttco) = 1.2 \times 10^{14}\left[\frac{(T_{\rm ex}+0.88)e^{5.3/T_{\rm ex}}}{1-e^{-10.6/T_{\rm ex}}}\right]\int \tau_{13}\,dv\;.
\end{equation}
A corresponding H$_2$ column density is derived using an abundance ratio of
\begin{equation}\label{eqn:abund13}
\Upsilon_{\rm 13CO} \equiv \frac{N(\rm H_2)}{N(\rm ^{13}CO)} = 3 \times 10^6\,,
\end{equation}
for consistency with the values inferred or adopted by previous analyses \citep{heikkila:99,mizuno:10,fujii:14}.

We also compute a luminosity-based H$_2$ mass directly from the \twco\ integrated intensity {by} assuming a constant CO-to-H$_2$ conversion factor:
\begin{equation}\label{eqn:xco}
X_{\rm CO} \equiv \frac{N(\rm H_2)}{I(\rm CO)} = 2 \times 10^{20}\,X_2\, \frac{\rm cm^{-2}}{\rm K\, \kms}\,.
\end{equation}
Here $X_2=1$ for a standard (Galactic) CO to H$_2$ conversion factor \citep{bolatto:13a}. In our analysis we assume $X_2=2.4$ for the CO(1--0) line (based on the virial analysis of the MAGMA GMC catalog by \citealt{hughes:10}) which translates to $X_2=1.6$ for the CO(2--1) line, adopting a CO(2--1)/CO(1--0) brightness temperature ratio of $R_{21} = 1.5$.
We adopt this value of $R_{21}$ based on a comparison of the ALMA TP spectra with resolution-matched MAGMA CO(1--0) spectra from \citet{wong:11}.
Previous work has shown the line ratio to vary with cloud conditions, with values $\sim$0.6 for molecular clouds in the outskirts of the LMC \citep{wong:17} and rising to $\sim$1 near 30 Dor (at 9\arcmin\ resolution, \citealt{sorai:01}), so a fixed value is only roughly appropriate.  While values of $R_{21} \gtrsim 1$ are not expected for optically thick, thermalized emission, they have been reported in other actively star-forming regions, in both Galactic \citep[Orion KL,][]{nishimura:15} and Magellanic (e.g.\ N83 in SMC, \citealt{bolatto:03}; N11 in LMC, \citealt{israel:03}) environments. As discussed by \citet{bolatto:03}, high $R_{21}$ can arise from a molecular medium that is both warm and clumpy (as is clearly the case for 30 Dor), since the larger photosphere ($\tau\sim 1$ surface) for the 2$\rightarrow$1 line fills more of the telescope beam.
Given the many uncertain assumptions in our analysis, and the likelihood that $X_{\rm CO}$ varies on scales comparable to or smaller than our map (see further discussion in \S\ref{sec:disc}), our luminosity-based masses should be considered uncertain by a factor of 2, and possibly more if substantial CO-dark gas is present.

\begin{figure*}
\includegraphics[width=\textwidth]{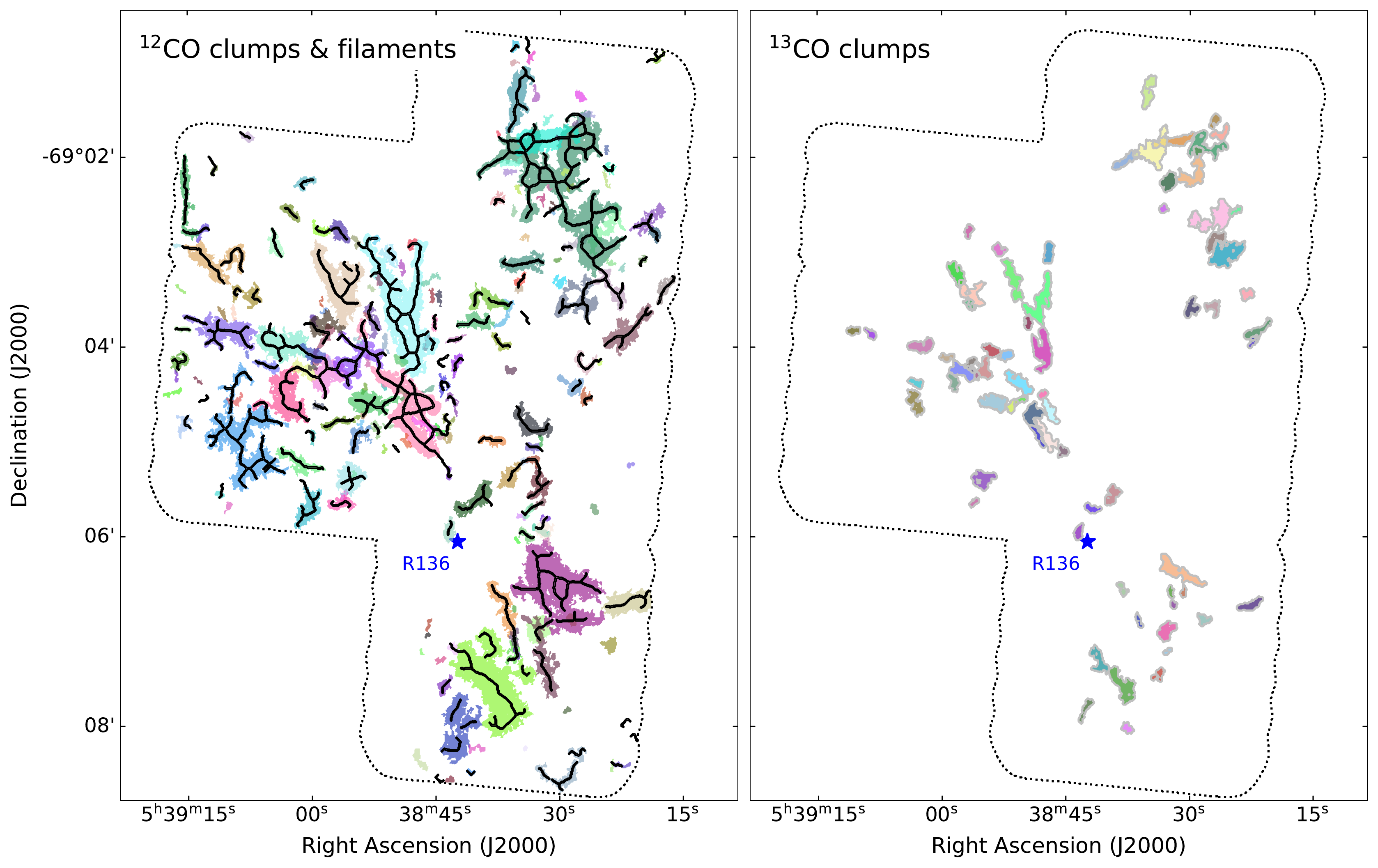}\\
\hspace*{0.6cm}
\includegraphics[width=0.95\textwidth]{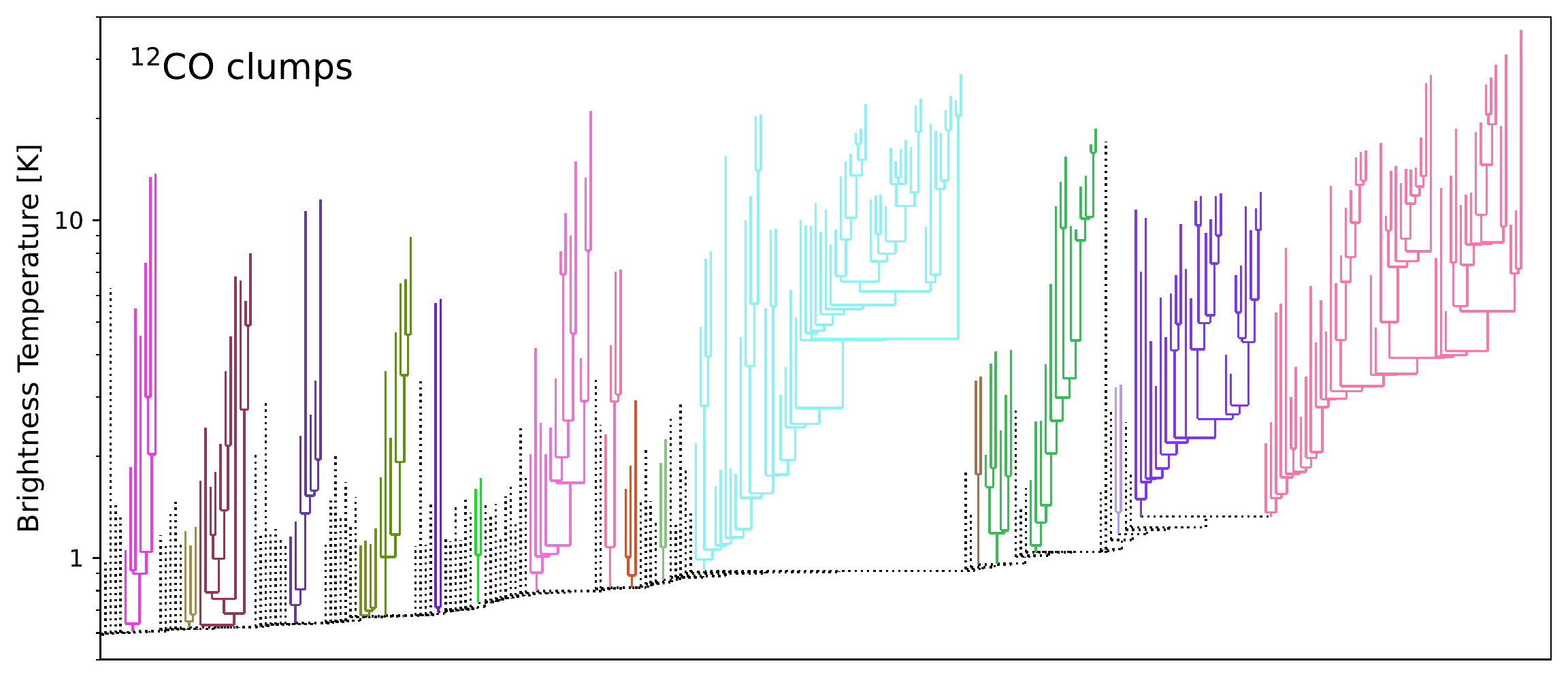}
\caption{Projected maps of the \twco\ ({top left}) and \ttco\ ({top right}) clumps identified by the SCIMES segmentation algorithm.  Each clump is shaded with a different color.  The filament skeleton identified by {\tt fil\_finder} is shown in black against the \twco\ clumps, but note that the filaments are identified in the CO peak SNR image whereas the clumps are identified in the cubes.  {The bottom panel shows a zoomed view of part of the dendrogram tree diagram for \twco\ emission, with clumps identified using the same colors as in the top left panel.  Dotted lines indicate dendrogram structures that are not identified as clumps by SCIMES.}
\label{fig:clust}}
\end{figure*}

\subsection{Structural Decomposition}\label{sec:dendro}

We use the Python program {\tt astrodendro}\footnote{\url{http://www.dendrograms.org}} to identify and segment the line emission regions in the cubes \citep{rosolowsky:08}. Parameters for the algorithm are chosen to identify local maxima in the cube above the 3$\sigma_{\rm rms}$ level that are also at least 2.5$\sigma_{\rm rms}$ above the merge level with adjacent structures.  Each local maximum is required to span at least two synthesized beams in area and is bounded by an isosurface at either the minimum (3$\sigma_{\rm rms}$) level or at the merge level with an adjoining structure.  Bounding isosurfaces surrounding the local maxima are categorized as {\it trunks}, {\it branches}, or {\it leaves} according to whether they are the largest contiguous structures (trunks), are intermediate in scale (branches), or have no resolved substructure (leaves).  Although the dendrogram structures are not all independent, trunks do not overlap other trunks {in the cube} and leaves do not overlap other leaves {in the cube}.  Since an object with no detected substructure is classified as a leaf, every trunk will contain leaf (and usually branch) substructures, which are collectively termed its {\it descendants}.

The basic properties of the identified structures are also determined by {\tt astrodendro}, including their spatial and velocity centroids ($\bar{x}, \bar{y}, \bar{v}$), the integrated flux $S$, rms line width $\sigma_v$ (defined as the intensity-weighted second moment of the structure along the velocity axis), the position angle of the major axis (as determined by principal component analysis) $\phi$, and the rms sizes along the major and minor axes, $\sigma_{\rm maj}$ and $\sigma_{\rm min}$. 
All properties are determined using the ``bijection'' approach discussed by \citet{rosolowsky:08}, which associates all emission bounded by an isosurface with the identified structure.
{We then calculate deconvolved values for the major and minor axes, $\sigma_{\rm maj}^\prime$ and $\sigma_{\rm min}^\prime$, approximating each structure as a 2-D Gaussian with major and minor axes of $\sigma_{\rm maj}$ and $\sigma_{\rm min}$ before deconvolving the telescope beam.  Structures which cannot be deconvolved are excluded from further analysis.}
From these basic properties we have calculated additional properties, including {the effective rms spatial size, $\sigma_r = \sqrt{\sigma_{\rm maj}^\prime \sigma_{\rm min}^\prime}$; the effective radius $R = 1.91 \sigma_r$,} following \citet{solomon:87}; the luminosity $L=Sd^2$, adopting $d=50$ kpc \citep{pietrzynski:19}; the virial mass $M_{\rm vir}=5\sigma_v^2R/G$, derived from solving the equilibrium condition (for kinetic energy ${\cal T}$ and potential energy ${\cal W}$):
\begin{equation}\label{eq:vireq}
2{\cal T} + {\cal W} = 2\left(\frac{3}{2}M_{\rm vir}\sigma_v^2\right) - \frac{3}{5}\frac{GM_{\rm vir}^2}{R} = 0\,;
\end{equation}
the LTE-based mass (from $^{13}$CO):
\begin{equation}
M_{\rm LTE} = (2m_p)(1.36)\Upsilon_{\rm 13CO} \int N(\ttco)\,dA \,,
\end{equation}
where the integration is over the projected area of the structure $A$, 1.36 is a correction factor for associated helium, and the abundance ratio $\Upsilon_{\rm 13CO}$ is given by Equation~\ref{eqn:abund13};
and the luminosity-based mass (from $^{12}$CO):
\begin{equation}
\frac{M_{\rm lum}}{M_\odot} = 4.3X_2\, \frac{L_{\rm CO}}{\rm K\;km\;s^{-1}\;pc^2}\,,
\end{equation}
where $X_2$ is defined in Equation~\ref{eqn:xco} and the factor of 4.3 includes associated helium \citep{bolatto:13a}.
{By taking ratios of these mass estimates we then calculate the so-called virial parameter,}
\begin{equation}
  \avir = \left\{
    \begin{array}{ll}
      M_{\rm vir}/M_{\rm lum} & \mbox{for \twco},\\
      M_{\rm vir}/M_{\rm LTE} & \mbox{for \ttco}.
    \end{array}
  \right.
\end{equation}
{Tables~\ref{tab:dendro12} and \ref{tab:dendro13} present the measured and derived properties of the resolved CO and \ttco\ dendrogram structures, including their classification as trunks, branches, or leaves.} 

We also post-process the dendrogram output using the SCIMES algorithm \citep{colombo:15}, which utilizes spectral clustering {(an unsupervised classification approach based on graph theory)} to identify discrete structures with similar emission properties.  The resulting clusters (hereafter referred to as {\it clumps} to avoid confusion with star clusters) form a set of independent objects, {avoiding the problem that the complete set of dendrogram structures constitute a nested rather than independent set.  At the same time, the SCIMES clumps span} a wider range of size, line width, and luminosity in comparison to the leaves, and because they are required to contain substructure, they are less likely to be influenced by fluctuations in the map noise.  In particular, we run the algorithm with the {\tt save\_branches} setting active, which retains isolated branches as clumps but not isolated leaves.  We use the ``volume'' criterion for defining similarity, which calculates volume as $V=\pi R^2\sigma_v$ for each structure.  Comparison runs using both ``volume'' and ``luminosity'' criteria, and without the {\tt save\_branches} setting, produce almost identical results for our data.  Note that because the clumps are a subset of the cataloged dendrogram structures, their properties have already been calculated as described above.  {Tables~\ref{tab:clust12} and \ref{tab:clust13} present the properties of the CO and \ttco\ clumps respectively, ordered by right ascension.}
Images of the individual \twco\ and \ttco\ clumps are shown in {the upper panels of} Figure~\ref{fig:clust}; since the clumps are identified in the cube, they are sometimes found projected against one another.  {The number of clumps found in \twco\ (\ttco) are 198 (71), of which 142 (61) have sizes which can be deconvolved.  The lower panel of Figure~\ref{fig:clust} shows a zoomed view of part of the \twco\ dendrogram tree, with the SCIMES clumps identified as distinctly colored sub-trees (the colors are chosen to match the upper left panel).}  We stress that the analyses of the \twco\ and \ttco\ data are conducted independently; we examine positional matches between the two sets of catalogs in \S\ref{sec:virial}.

\subsection{Filament Identification}\label{sec:filfinder}

We also employed an alternative structure-finding package, FilFinder, to highlight the filamentary nature of the emission.  We apply the {\tt FilFinder2D} algorithm, described in \citet{koch:15}, to the peak SNR image of \twco(2--1) emission.  To suppress bright regions, the image is first flattened with an arctan transform, $I^\prime = I_0 \arctan(I/I_0)$, where $I_0$ is chosen as the 80th percentile of the image brightness distribution (for this image $I_0 = 5.3\sigma_{\rm rms}$).  A mask is then created from the flattened image using adaptive thresholding with the following parameters: {\tt smooth\_size} of 5 pixels (corresponding to 2\farcs5), {\tt adapt\_thresh} of 10 pixels (corresponding to 5\arcsec), {\tt size\_thresh} of 80 pixels (corresponding to 20 arcsec$^2$), and {\tt glob\_thresh} of 4$\sigma$.  We experimented with a variety of parameter sets but found that these parameters produced a signal mask that was most consistent with the emission regions identified with SCIMES.  Each mask region is reduced to a one-pixel wide ``skeleton'' using the medial axis transform, and small structures are removed by imposing a minimum length (pixel count) of 4 beam widths for the skeleton as a whole and 2 beam widths for branches that depart from the longest path through the skeleton.  The resulting skeletonization of the emission, after pruning of small structures, is visualized in black {in the upper left} panel of Figure~\ref{fig:clust}.  The skeletonization is effective at identifying and connecting large, coherent emission structures, but ``breaks'' in the filamentary structure may still arise from sensitivity limitations that prevent the algorithm from connecting neighboring skeletons.
{While it is possible that velocity discontinuities across filaments could be missed by identifying filaments only in 2-D, we generally observe that spatially coherent filaments are also coherent in velocity.}

\begin{deluxetable*}{ccccDccRrR@{$\pm$}cR@{$\pm$}cR@{$\pm$}cR@{$\pm$}cR@{$\pm$}crc}
\rotate
\tablecaption{All Resolved Structures in the Default \twco\ ALMA 30 Dor Cube\label{tab:dendro12}}
\tablehead{
\colhead{No.} & \colhead{R. A.} & \colhead{Decl.} & \colhead{$v_{\rm LSR}$} & \twocolhead{CO Flux} & \colhead{$\sigma_{\rm maj}$} & \colhead{$\sigma_{\rm min}$} & \colhead{$\phi$\tablenotemark{a}} & \colhead{$A$\tablenotemark{b}} & \twocolhead{$\log\, R$} & \twocolhead{$\log\, \sigma_v$} & \twocolhead{$\log\, M_{\rm lum}$} & \twocolhead{$\log\, M_{\rm vir}$} & \twocolhead{$\log\, \alpha_{\rm vir}$} & \colhead{$\theta_{\rm off}$} & \colhead{Type\tablenotemark{c}}\\
& \colhead{(J2000)} & \colhead{(J2000)} & \colhead{(\kms)} & \twocolhead{(Jy \kms)} & \colhead{($\mathrm{{}^{\prime\prime}}$)} & \colhead{($\mathrm{{}^{\prime\prime}}$)} & \colhead{($\mathrm{{}^{\circ}}$)} & \colhead{(pc$^2$)} & \twocolhead{(pc)} & \twocolhead{(\kms)} & \twocolhead{($M_\odot$)} & \twocolhead{($M_\odot$)} & \twocolhead{} & \colhead{($\mathrm{{}^{\prime\prime}}$)}}
\decimals
\startdata
1 & 05:38:17.24 & -69:03:23.0 & 250.31 & 15.73 & 3.22 & 0.98 & 48 & 2.48 & -0.18 & 0.05 & -0.07 & 0.04 & 2.22 & 0.04 & 2.75 & 0.08 & 0.53 & 0.09 & 209 & B \\
2 & 05:38:17.36 & -69:03:24.1 & 249.98 & 29.62 & 4.78 & 1.33 & 58 & 6.26 & 0.02 & 0.04 & 0.11 & 0.04 & 2.49 & 0.04 & 3.31 & 0.08 & 0.82 & 0.09 & 208 & B \\
3 & 05:38:17.37 & -69:03:24.1 & 250.03 & 30.07 & 4.77 & 1.39 & 58 & 6.54 & 0.04 & 0.04 & 0.13 & 0.04 & 2.50 & 0.04 & 3.37 & 0.08 & 0.87 & 0.09 & 208 & T \\
4 & 05:38:17.45 & -69:03:24.3 & 250.40 & 8.56 & 1.37 & 1.01 & 64 & 1.25 & -0.38 & 0.06 & -0.15 & 0.05 & 1.95 & 0.04 & 2.38 & 0.09 & 0.43 & 0.10 & 207 & L \\
5 & 05:38:17.92 & -69:02:32.8 & 260.89 & 3.94 & 1.89 & 1.10 & 92 & 2.35 & -0.26 & 0.06 & 0.03 & 0.05 & 1.62 & 0.04 & 2.87 & 0.09 & 1.25 & 0.10 & 248 & B \\
6 & 05:38:17.93 & -69:02:32.3 & 260.80 & 3.06 & 1.36 & 1.00 & 86 & 1.45 & -0.39 & 0.07 & 0.03 & 0.05 & 1.51 & 0.04 & 2.73 & 0.11 & 1.22 & 0.12 & 248 & L \\
7 & 05:38:18.24 & -69:00:58.0 & 260.20 & 5.55 & 2.03 & 0.95 & 46 & 1.87 & -0.31 & 0.08 & -0.32 & 0.06 & 1.77 & 0.04 & 2.11 & 0.12 & 0.34 & 0.12 & 331 & T \\
8 & 05:38:18.32 & -69:00:58.4 & 260.15 & 4.05 & 1.11 & 0.94 & 53 & 1.20 & -0.49 & 0.10 & -0.38 & 0.07 & 1.63 & 0.04 & 1.82 & 0.14 & 0.19 & 0.15 & 331 & L \\
9 & 05:38:18.48 & -69:02:47.7 & 253.97 & 5.14 & 1.92 & 1.01 & 144 & 1.95 & -0.29 & 0.05 & -0.19 & 0.05 & 1.73 & 0.04 & 2.39 & 0.09 & 0.65 & 0.10 & 234 & B \\
10 & 05:38:18.49 & -69:02:48.1 & 253.94 & 6.80 & 2.20 & 1.36 & 127 & 3.29 & -0.15 & 0.04 & -0.12 & 0.04 & 1.86 & 0.04 & 2.68 & 0.08 & 0.83 & 0.09 & 233 & T \\
\enddata
\tablenotetext{a}{Position angle is measured counterclockwise from $+x$ direction (west).}
\tablenotetext{b}{Projected area of clump.}
\tablenotetext{c}{Type of structure: (T)runk, (B)ranch, or (L)eaf.}
\tablecomments{Table~\ref{tab:dendro12} is published in its entirety in machine-readable format.
A portion is shown here for guidance regarding its form and content.}
\end{deluxetable*}

\begin{deluxetable*}{ccccDccRrR@{$\pm$}cR@{$\pm$}cR@{$\pm$}cR@{$\pm$}cR@{$\pm$}crc}
\rotate
\tablecaption{All Resolved Structures in the Default \ttco\ ALMA 30 Dor Cube\label{tab:dendro13}}
\tablehead{
\colhead{No.} & \colhead{R. A.} & \colhead{Decl.} & \colhead{$v_{\rm LSR}$} & \twocolhead{\ttco\ Flux} & \colhead{$\sigma_{\rm maj}$} & \colhead{$\sigma_{\rm min}$} & \colhead{$\phi$\tablenotemark{a}} & \colhead{$A$\tablenotemark{b}} & \twocolhead{$\log\, R$} & \twocolhead{$\log\, \sigma_v$} & \twocolhead{$\log\, M_{\rm LTE}$} & \twocolhead{$\log\, M_{\rm vir}$} & \twocolhead{$\log\, \alpha_{\rm vir}$} & \colhead{$\theta_{\rm off}$} & \colhead{Type\tablenotemark{c}}\\
& \colhead{(J2000)} & \colhead{(J2000)} & \colhead{(\kms)} & \twocolhead{(Jy \kms)} & \colhead{($\mathrm{{}^{\prime\prime}}$)} & \colhead{($\mathrm{{}^{\prime\prime}}$)} & \colhead{($\mathrm{{}^{\circ}}$)} & \colhead{(pc$^2$)} & \twocolhead{(pc)} & \twocolhead{(\kms)} & \twocolhead{($M_\odot$)} & \twocolhead{($M_\odot$)} & \twocolhead{} & \colhead{($\mathrm{{}^{\prime\prime}}$)}}
\decimals
\startdata
1 & 05:38:19.81 & -69:06:41.5 & 255.44 & 7.58 & 1.35 & 0.96 & 94 & 1.97 & -0.41 & 0.08 & -0.14 & 0.06 & 2.74 & 0.04 & 2.38 & 0.12 & -0.37 & 0.13 & 126 & L \\
2 & 05:38:22.44 & -69:03:51.5 & 253.36 & 10.88 & 3.78 & 1.04 & -153 & 5.04 & -0.12 & 0.05 & 0.14 & 0.04 & 2.85 & 0.04 & 3.23 & 0.08 & 0.37 & 0.09 & 169 & T \\
3 & 05:38:22.49 & -69:06:43.6 & 254.62 & 4.56 & 2.42 & 0.88 & -157 & 2.14 & -0.32 & 0.14 & -0.51 & 0.07 & 2.52 & 0.04 & 1.72 & 0.17 & -0.80 & 0.17 & 113 & L \\
4 & 05:38:22.60 & -69:06:43.6 & 254.62 & 6.98 & 3.47 & 1.09 & -161 & 4.07 & -0.12 & 0.04 & -0.44 & 0.04 & 2.69 & 0.04 & 2.06 & 0.08 & -0.62 & 0.09 & 113 & T \\
5 & 05:38:22.68 & -69:03:52.1 & 253.26 & 7.07 & 2.12 & 0.78 & -160 & 2.00 & -0.49 & 0.23 & 0.11 & 0.04 & 2.67 & 0.04 & 2.80 & 0.24 & 0.13 & 0.24 & 168 & B \\
6 & 05:38:22.80 & -69:03:26.4 & 252.46 & 1.60 & 1.14 & 0.95 & 68 & 1.32 & -0.48 & 0.12 & -0.09 & 0.06 & 2.01 & 0.04 & 2.40 & 0.14 & 0.39 & 0.15 & 189 & L \\
7 & 05:38:23.16 & -69:03:26.8 & 250.56 & 4.04 & 1.91 & 1.36 & -171 & 2.57 & -0.18 & 0.04 & 0.26 & 0.04 & 2.41 & 0.04 & 3.39 & 0.08 & 0.98 & 0.09 & 187 & T \\
8 & 05:38:24.56 & -69:03:01.4 & 251.14 & 7.08 & 2.08 & 0.94 & -140 & 1.88 & -0.31 & 0.06 & -0.03 & 0.04 & 2.69 & 0.04 & 2.69 & 0.09 & -0.00 & 0.10 & 205 & L \\
9 & 05:38:25.81 & -69:03:02.5 & 252.03 & 55.22 & 4.81 & 2.61 & -157 & 13.94 & 0.20 & 0.04 & 0.18 & 0.04 & 3.59 & 0.04 & 3.63 & 0.08 & 0.04 & 0.09 & 201 & B \\
10 & 05:38:25.83 & -69:06:33.9 & 249.77 & 1.74 & 1.80 & 0.95 & 157 & 1.94 & -0.34 & 0.11 & -0.35 & 0.09 & 2.07 & 0.04 & 2.02 & 0.17 & -0.05 & 0.18 & 93 & L \\
\enddata
\tablenotetext{a}{Position angle is measured counterclockwise from $+x$ direction (west).}
\tablenotetext{b}{Projected area of clump.}
\tablenotetext{c}{Type of structure: (T)runk, (B)ranch, or (L)eaf.}
\tablecomments{Table~\ref{tab:dendro13} is published in its entirety in machine-readable format.
A portion is shown here for guidance regarding its form and content.}
\end{deluxetable*}

\begin{deluxetable*}{ccccDccRrR@{$\pm$}cR@{$\pm$}cR@{$\pm$}cR@{$\pm$}cR@{$\pm$}cr}
\rotate
\tablecaption{SCIMES Clumps in the Default \twco\ ALMA 30 Dor Cube\label{tab:clust12}}
\tablehead{
\colhead{No.} & \colhead{R. A.} & \colhead{Decl.} & \colhead{$v_{\rm LSR}$} & \twocolhead{CO Flux} & \colhead{$\sigma_{\rm maj}$} & \colhead{$\sigma_{\rm min}$} & \colhead{$\phi$\tablenotemark{a}} & \colhead{$A$\tablenotemark{b}} & \twocolhead{$\log\, R$} & \twocolhead{$\log\, \sigma_v$} & \twocolhead{$\log\, M_{\rm lum}$} & \twocolhead{$\log\, M_{\rm vir}$} & \twocolhead{$\log\, \alpha_{\rm vir}$} & \colhead{$\theta_{\rm off}$}\\
& \colhead{(J2000)} & \colhead{(J2000)} & \colhead{(\kms)} & \twocolhead{(Jy \kms)} & \colhead{($\mathrm{{}^{\prime\prime}}$)} & \colhead{($\mathrm{{}^{\prime\prime}}$)} & \colhead{($\mathrm{{}^{\circ}}$)} & \colhead{(pc$^2$)} & \twocolhead{(pc)} & \twocolhead{(\kms)} & \twocolhead{($M_\odot$)} & \twocolhead{($M_\odot$)} & \twocolhead{} & \colhead{($\mathrm{{}^{\prime\prime}}$)}}
\decimals
\startdata
1 & 05:38:17.37 & -69:03:24.1 & 250.03 & 30.07 & 4.77 & 1.39 & 58 & 6.54 & 0.04 & 0.04 & 0.13 & 0.04 & 2.50 & 0.04 & 3.37 & 0.08 & 0.87 & 0.09 & 208 \\
2 & 05:38:18.24 & -69:00:58.0 & 260.20 & 5.55 & 2.03 & 0.95 & 46 & 1.87 & -0.31 & 0.08 & -0.32 & 0.06 & 1.77 & 0.04 & 2.11 & 0.12 & 0.34 & 0.12 & 331 \\
3 & 05:38:18.49 & -69:02:48.1 & 253.94 & 6.80 & 2.20 & 1.36 & 127 & 3.29 & -0.15 & 0.04 & -0.12 & 0.04 & 1.86 & 0.04 & 2.68 & 0.08 & 0.83 & 0.09 & 233 \\
4 & 05:38:19.41 & -69:02:39.6 & 260.03 & 18.58 & 6.60 & 1.85 & -140 & 10.36 & 0.19 & 0.04 & 0.19 & 0.04 & 2.29 & 0.04 & 3.64 & 0.08 & 1.35 & 0.09 & 238 \\
5 & 05:38:20.12 & -69:03:05.4 & 258.60 & 3.83 & 1.47 & 0.89 & 45 & 1.67 & -0.44 & 0.08 & 0.19 & 0.05 & 1.61 & 0.04 & 3.00 & 0.11 & 1.39 & 0.12 & 214 \\
6 & 05:38:21.56 & -69:06:42.4 & 254.71 & 160.67 & 8.18 & 2.36 & -170 & 22.34 & 0.30 & 0.04 & -0.03 & 0.04 & 3.23 & 0.04 & 3.30 & 0.08 & 0.07 & 0.09 & 118 \\
7 & 05:38:22.07 & -69:03:51.4 & 252.69 & 245.96 & 6.85 & 2.65 & -139 & 22.99 & 0.28 & 0.04 & 0.39 & 0.04 & 3.41 & 0.04 & 4.12 & 0.08 & 0.71 & 0.09 & 171 \\
8 & 05:38:22.41 & -69:08:24.6 & 247.02 & 2.49 & 3.11 & 0.77 & -179 & 1.53 & -0.45 & 0.46 & -0.10 & 0.08 & 1.42 & 0.04 & 2.43 & 0.48 & 1.01 & 0.48 & 177 \\
9 & 05:38:22.54 & -69:03:10.0 & 252.45 & 0.54 & 1.24 & 0.77 & 62 & 0.90 & -0.70 & 0.47 & -0.06 & 0.09 & 0.75 & 0.04 & 2.24 & 0.49 & 1.49 & 0.49 & 203 \\
10 & 05:38:23.48 & -69:03:25.2 & 251.12 & 116.85 & 6.80 & 2.46 & 127 & 11.68 & 0.27 & 0.04 & 0.30 & 0.04 & 3.09 & 0.04 & 3.93 & 0.08 & 0.84 & 0.09 & 188 \\
\enddata
\tablenotetext{a}{Position angle is measured counterclockwise from $+x$ direction (west).}
\tablenotetext{b}{Projected area of clump.}
\tablecomments{Table~\ref{tab:clust12} is published in its entirety in machine-readable format.
A portion is shown here for guidance regarding its form and content.}
\end{deluxetable*}

\begin{deluxetable*}{ccccDccRrR@{$\pm$}cR@{$\pm$}cR@{$\pm$}cR@{$\pm$}cR@{$\pm$}cr}
\rotate
\tablecaption{SCIMES Clumps in the Default \ttco\ ALMA 30 Dor Cube\label{tab:clust13}}
\tablehead{
\colhead{No.} & \colhead{R. A.} & \colhead{Decl.} & \colhead{$v_{\rm LSR}$} & \twocolhead{\ttco\ Flux} & \colhead{$\sigma_{\rm maj}$} & \colhead{$\sigma_{\rm min}$} & \colhead{$\phi$\tablenotemark{a}} & \colhead{$A$\tablenotemark{b}} & \twocolhead{$\log\, R$} & \twocolhead{$\log\, \sigma_v$} & \twocolhead{$\log\, M_{\rm LTE}$} & \twocolhead{$\log\, M_{\rm vir}$} & \twocolhead{$\log\, \alpha_{\rm vir}$} & \colhead{$\theta_{\rm off}$}\\
& \colhead{(J2000)} & \colhead{(J2000)} & \colhead{(\kms)} & \twocolhead{(Jy \kms)} & \colhead{($\mathrm{{}^{\prime\prime}}$)} & \colhead{($\mathrm{{}^{\prime\prime}}$)} & \colhead{($\mathrm{{}^{\circ}}$)} & \colhead{(pc$^2$)} & \twocolhead{(pc)} & \twocolhead{(\kms)} & \twocolhead{($M_\odot$)} & \twocolhead{($M_\odot$)} & \twocolhead{} & \colhead{($\mathrm{{}^{\prime\prime}}$)}}
\decimals
\startdata
1 & 05:38:22.44 & -69:03:51.5 & 253.36 & 10.88 & 3.78 & 1.04 & -153 & 5.04 & -0.12 & 0.05 & 0.14 & 0.04 & 2.85 & 0.04 & 3.23 & 0.08 & 0.37 & 0.09 & 169 \\
2 & 05:38:22.60 & -69:06:43.6 & 254.62 & 6.98 & 3.47 & 1.09 & -161 & 4.07 & -0.12 & 0.04 & -0.44 & 0.04 & 2.69 & 0.04 & 2.06 & 0.08 & -0.62 & 0.09 & 113 \\
3 & 05:38:23.16 & -69:03:26.8 & 250.56 & 4.04 & 1.91 & 1.36 & -171 & 2.57 & -0.18 & 0.04 & 0.26 & 0.04 & 2.41 & 0.04 & 3.39 & 0.08 & 0.98 & 0.09 & 187 \\
4 & 05:38:25.81 & -69:03:02.5 & 252.03 & 55.22 & 4.81 & 2.61 & -157 & 13.94 & 0.20 & 0.04 & 0.18 & 0.04 & 3.59 & 0.04 & 3.63 & 0.08 & 0.04 & 0.09 & 201 \\
5 & 05:38:26.30 & -69:01:45.6 & 247.89 & 3.65 & 3.16 & 1.57 & -154 & 3.13 & -0.02 & 0.04 & -0.10 & 0.04 & 2.37 & 0.04 & 2.84 & 0.08 & 0.47 & 0.09 & 272 \\
6 & 05:38:26.90 & -69:01:36.3 & 246.08 & 2.51 & 1.94 & 1.11 & 55 & 2.12 & -0.25 & 0.06 & -0.32 & 0.05 & 2.23 & 0.04 & 2.18 & 0.10 & -0.04 & 0.10 & 279 \\
7 & 05:38:27.11 & -69:02:38.5 & 250.61 & 31.76 & 7.85 & 3.31 & -164 & 14.23 & 0.37 & 0.04 & 0.08 & 0.04 & 3.33 & 0.04 & 3.59 & 0.08 & 0.26 & 0.09 & 220 \\
8 & 05:38:27.18 & -69:02:53.8 & 253.33 & 12.55 & 3.89 & 1.74 & -137 & 4.79 & 0.06 & 0.04 & 0.02 & 0.04 & 2.93 & 0.04 & 3.16 & 0.08 & 0.23 & 0.09 & 206 \\
9 & 05:38:27.27 & -69:03:34.9 & 253.35 & 3.02 & 2.32 & 1.80 & 155 & 2.95 & -0.06 & 0.04 & -0.08 & 0.04 & 2.29 & 0.04 & 2.85 & 0.08 & 0.56 & 0.09 & 169 \\
10 & 05:38:28.25 & -69:06:52.5 & 249.67 & 3.11 & 1.57 & 1.17 & 166 & 2.23 & -0.29 & 0.06 & -0.11 & 0.05 & 2.31 & 0.04 & 2.56 & 0.09 & 0.25 & 0.10 & 90 \\
\enddata
\tablenotetext{a}{Position angle is measured counterclockwise from $+x$ direction (west).}
\tablenotetext{b}{Projected area of clump.}
\tablecomments{Table~\ref{tab:clust13} is published in its entirety in machine-readable format.
A portion is shown here for guidance regarding its form and content.}
\end{deluxetable*}

\begin{deluxetable*}{cclrD@{ $\pm$}DD@{ $\pm$}Drr}
\tablehead{
\colhead{$Y$} & \colhead{$X$} & \colhead{Data Set} & \colhead{Number} & \multicolumn{4}{c}{$a_1$} & \multicolumn{4}{c}{$a_0$} & \colhead{$\chi^2_\nu$} & \colhead{$\varepsilon$\tablenotemark{a}}}
\tablecaption{Default Cubes --- Power Law Fit Parameters: $\log Y = a_1 \log X + a_0$\label{tab:fitpar}}
\decimals
\startdata
$\sigma_v$ & $R$ & \twco\ dendros & 1434 & 0.47 & 0.01 & 0.08 & 0.01 & 14.3 & 0.21\\
$\sigma_v$ & $R$ & \twco\ clumps & 142 & 0.47 & 0.06 & 0.13 & 0.02 & 14.3 & 0.21\\
$\sigma_v$ & $R$ & \ttco\ dendros & 254 & 0.73 & 0.06 & 0.06 & 0.01 & 10.5 & 0.22\\
$\sigma_v$ & $R$ & \ttco\ clumps & 61 & 1.42 & 0.37 & 0.06 & 0.04 & 14.3 & 0.35\\
$\Sigma_{\rm vir}$ & $\Sigma_{\rm lum}$ & \twco\ dendros & 1434 & 0.51 & 0.02 & 1.58 & 0.04 & 13.7 & 0.35\\
$\Sigma_{\rm vir}$ & $\Sigma_{\rm lum}$ & \twco\ clumps & 142 & 0.41 & 0.07 & 1.93 & 0.12 & 15.6 & 0.35\\
$\Sigma_{\rm vir}$ & $\Sigma_{\rm LTE}$ & \ttco\ dendros & 254 & 0.66 & 0.06 & 0.90 & 0.14 & 11.0 & 0.36\\
$\Sigma_{\rm vir}$ & $\Sigma_{\rm LTE}$ & \ttco\ clumps & 61 & 0.85 & 0.14 & 0.55 & 0.31 & 11.0 & 0.30\\
\enddata
\tablenotetext{a}{r.m.s.\ scatter in $\log Y$ relative to the best-fit line.  Units are dex.}
\end{deluxetable*}

\begin{deluxetable*}{cclrD@{ $\pm$}DD@{ $\pm$}Drr}
\tablehead{
\colhead{$Y$} & \colhead{$X$} & \colhead{Data Set} & \colhead{Number} & \multicolumn{4}{c}{$a_1$} & \multicolumn{4}{c}{$a_0$} & \colhead{$\chi^2_\nu$} & \colhead{$\varepsilon$\tablenotemark{a}}}
\tablecaption{0.1 \kms\ Cubes --- Power Law Fit Parameters: $\log Y = a_1 \log X + a_0$\label{tab:fitpar2}}
\decimals
\startdata
$\sigma_v$ & $R$ & \twco\ dendros & 2053 & 0.51 & 0.01 & 0.04 & 0.01 & 15.1 & 0.24\\
$\sigma_v$ & $R$ & \twco\ clumps & 221 & 0.76 & 0.06 & 0.09 & 0.02 & 13.6 & 0.28\\
$\sigma_v$ & $R$ & \ttco\ dendros & 310 & 0.74 & 0.05 & 0.06 & 0.01 & 13.2 & 0.24\\
$\sigma_v$ & $R$ & \ttco\ clumps & 72 & 0.91 & 0.17 & 0.09 & 0.03 & 13.5 & 0.28 \\
$\Sigma_{\rm vir}$ & $\Sigma_{\rm lum}$ & \twco\ dendros & 2053 & 0.57 & 0.01 & 1.43 & 0.03 & 12.9 & 0.34\\
$\Sigma_{\rm vir}$ & $\Sigma_{\rm lum}$ & \twco\ clumps & 221 & 0.55 & 0.04 & 1.64 & 0.07 & 11.8 & 0.33\\
$\Sigma_{\rm vir}$ & $\Sigma_{\rm LTE}$ & \ttco\ dendros & 310 & 0.79 & 0.05 & 0.56 & 0.12 & 11.8 & 0.34 \\
$\Sigma_{\rm vir}$ & $\Sigma_{\rm LTE}$ & \ttco\ clumps & 72 & 0.83 & 0.12 & 0.58 & 0.25 & 11.1 & 0.32\\
\enddata
\tablenotetext{a}{r.m.s.\ scatter in $\log Y$ relative to the best-fit line.  Units are dex.}
\end{deluxetable*}

\begin{figure*}[t]
\includegraphics[height=2.5in]{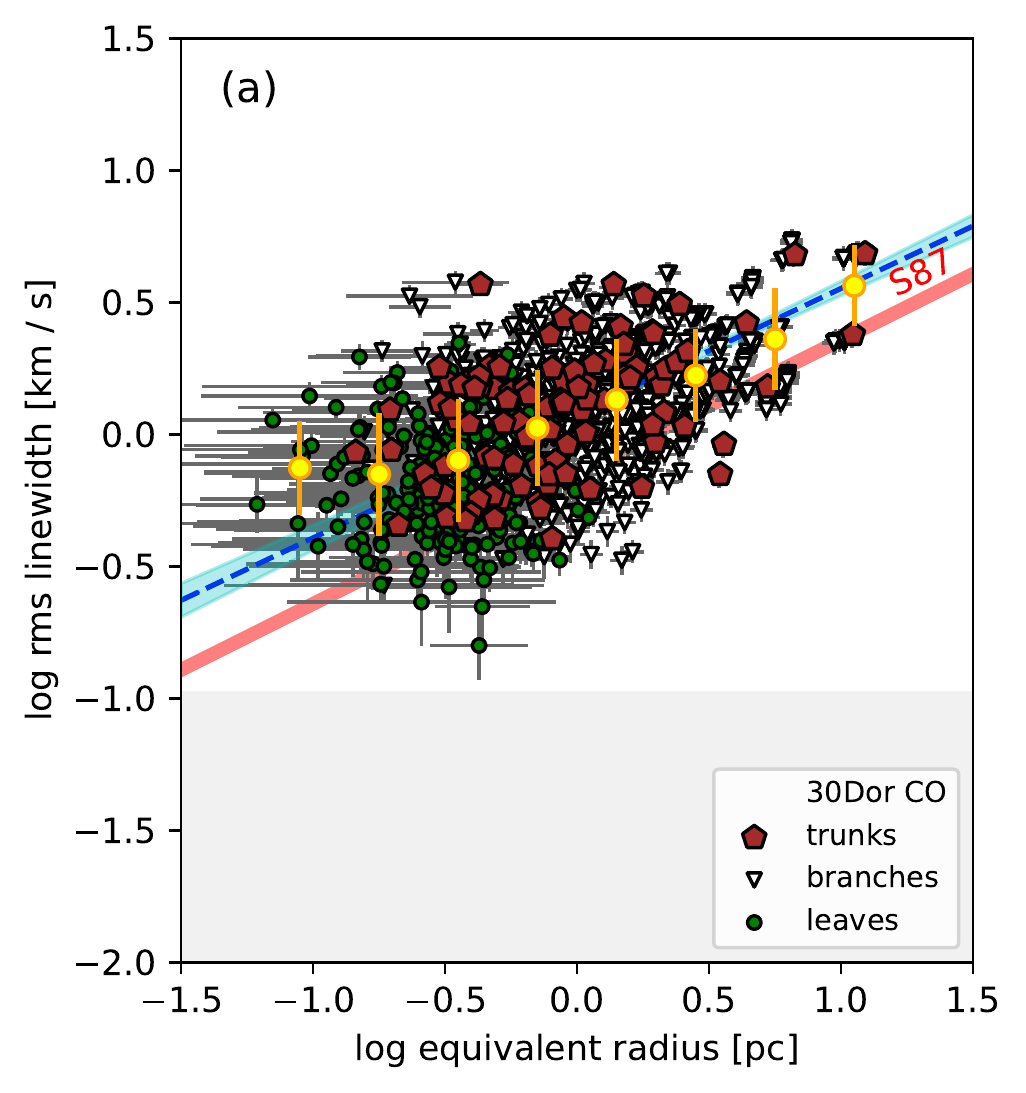}
\includegraphics[height=2.5in]{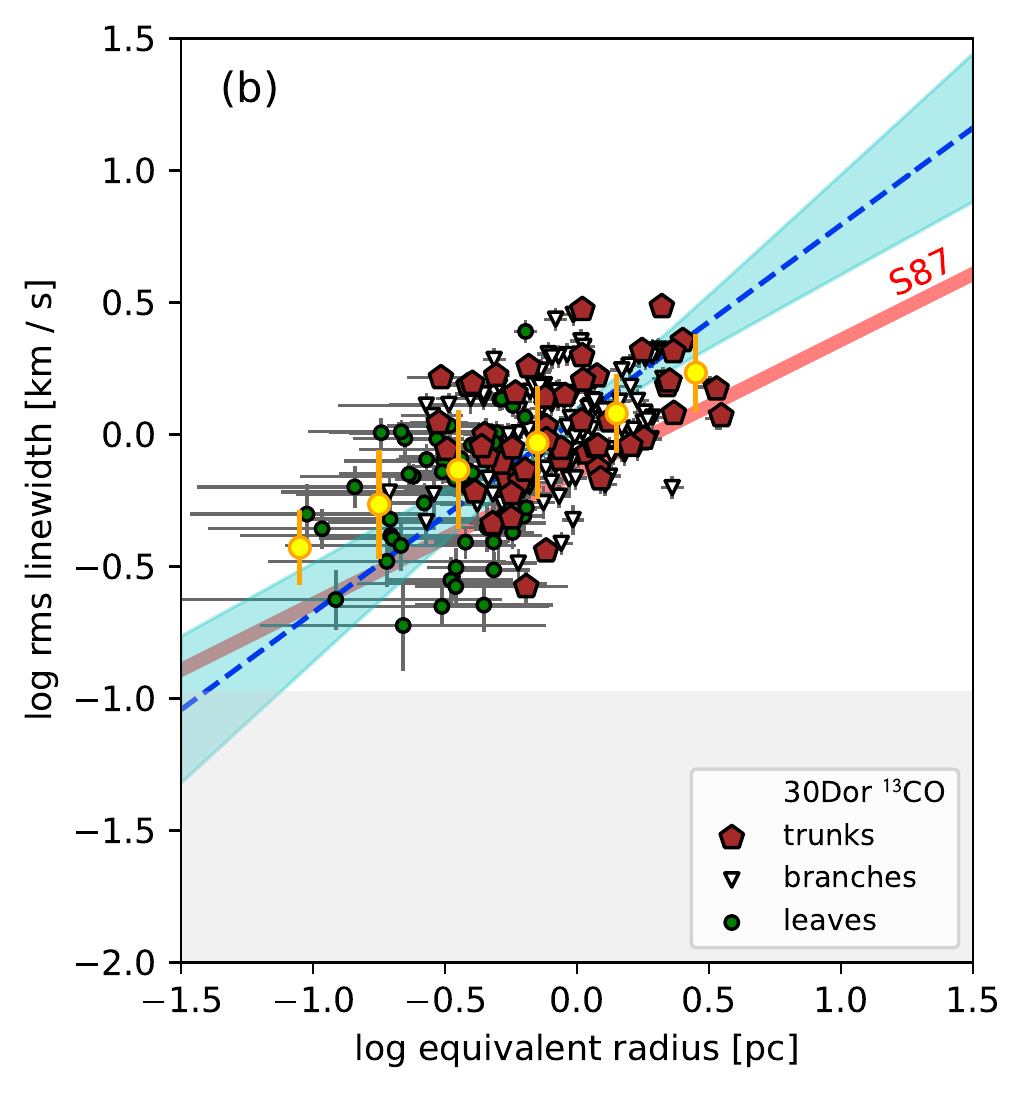}
\includegraphics[height=2.5in]{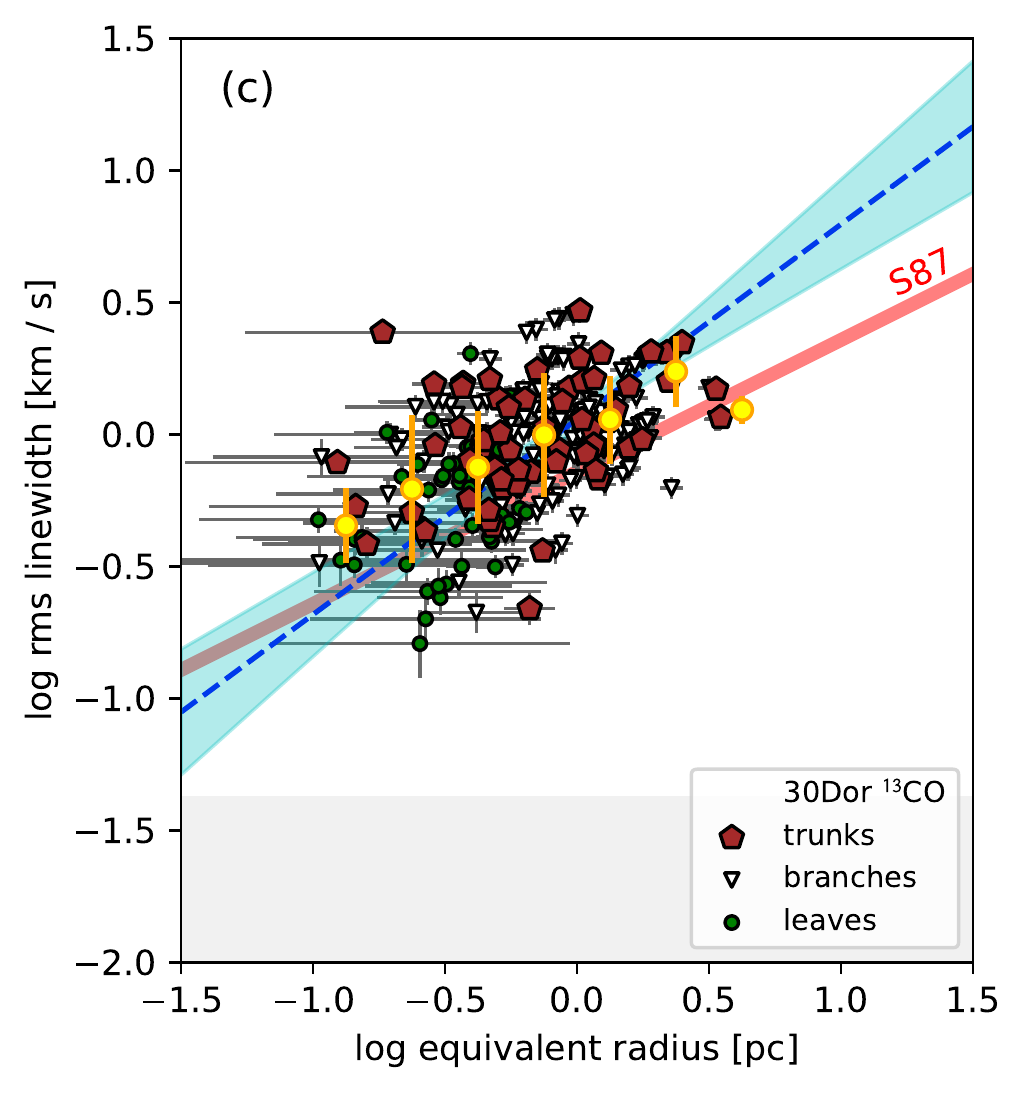}
\caption{Size-linewidth relations for dendrogram structures identified in the feathered data: (a) \twco\ structures; (b) \ttco\ structures; (c) \ttco\ structures at 0.1 \kms\ velocity resolution.  Different plot symbols distinguish the trunks, branches, and leaves of the dendrogram.  The power law fit and 3$\sigma$ uncertainty are shown in blue; {the gray shaded region indicates the limiting spectral resolution}.  Fit parameters are tabulated in Tables~\ref{tab:fitpar} and \ref{tab:fitpar2}.  Yellow circles are binned averages of all points.
\label{fig:rdv_feather}}
\end{figure*}

\begin{figure*}
\includegraphics[height=2.5in]{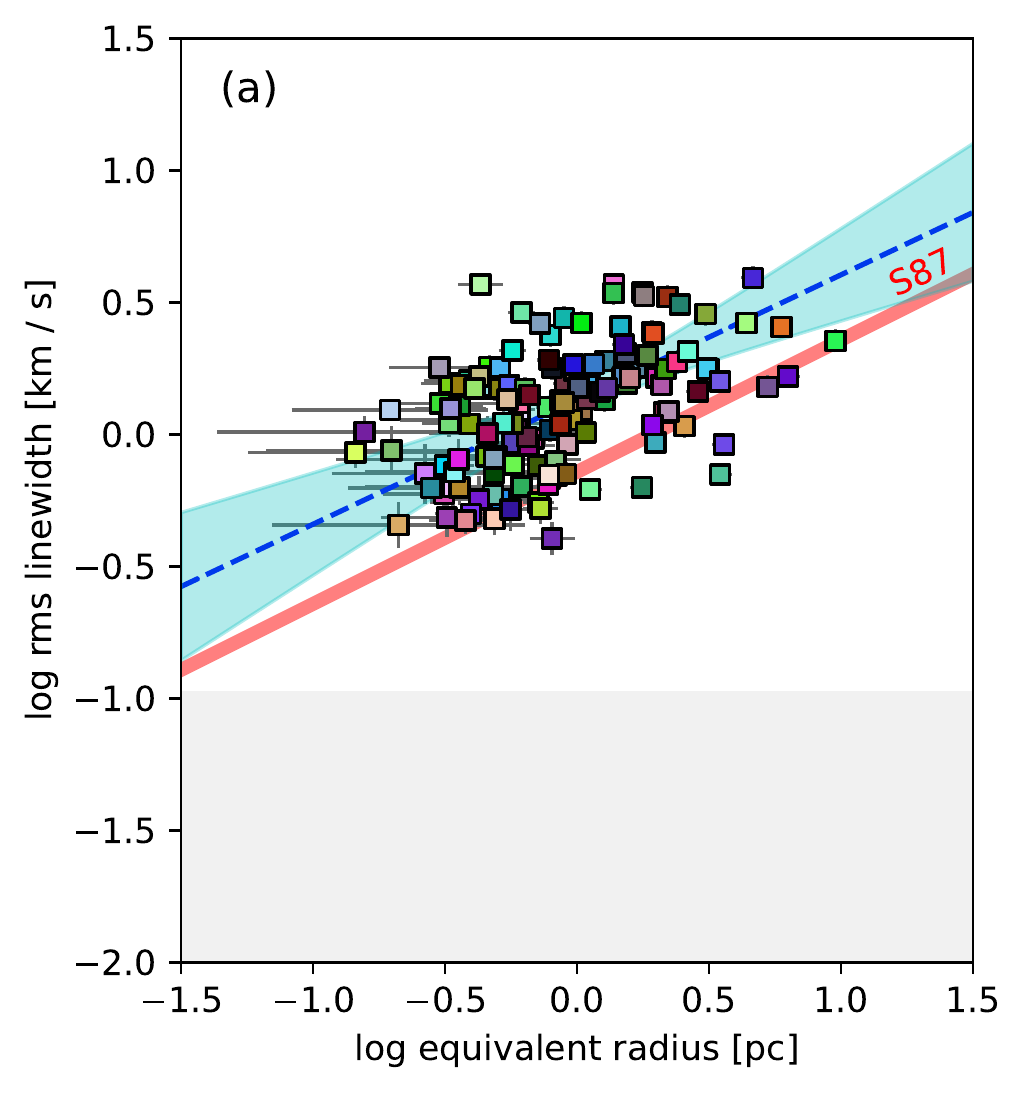}
\includegraphics[height=2.5in]{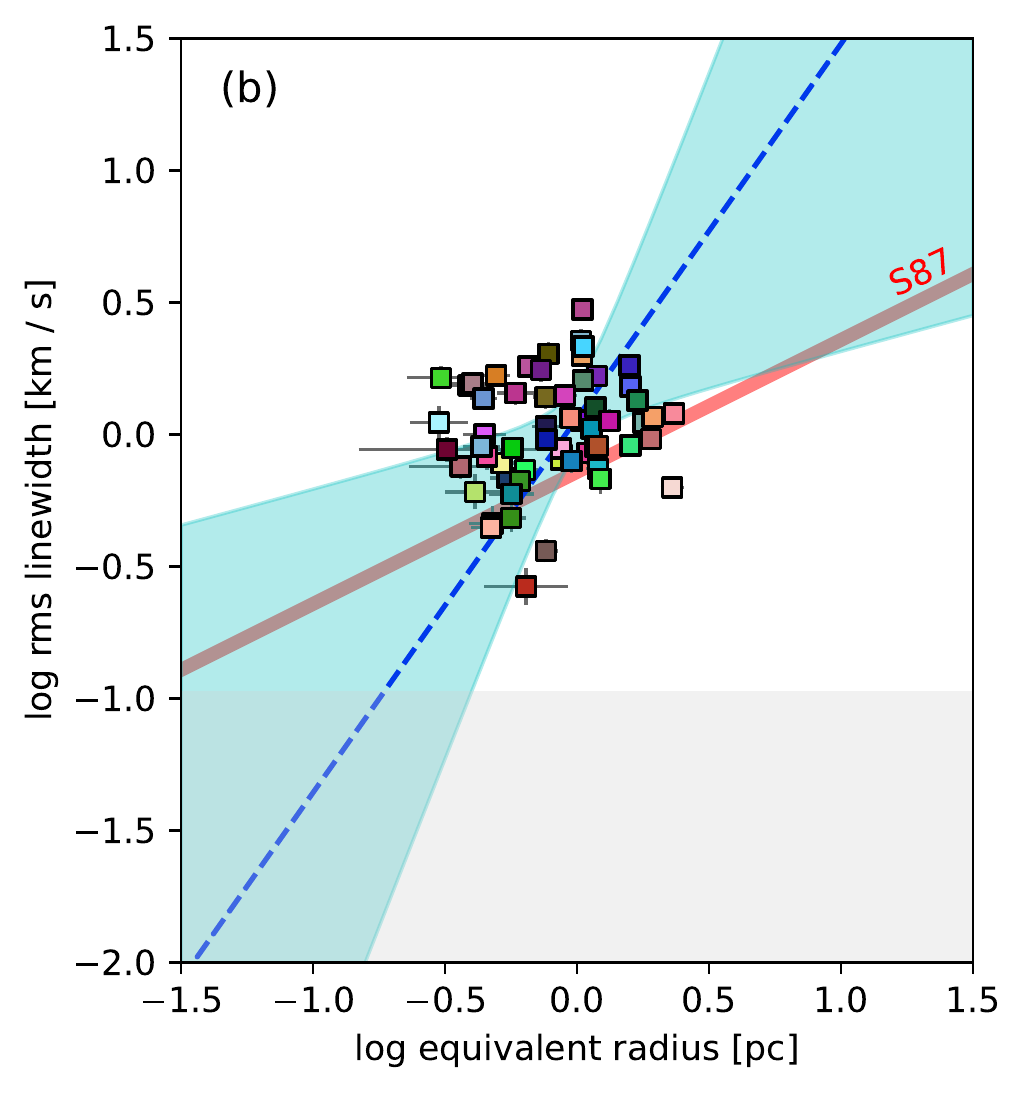}
\includegraphics[height=2.5in]{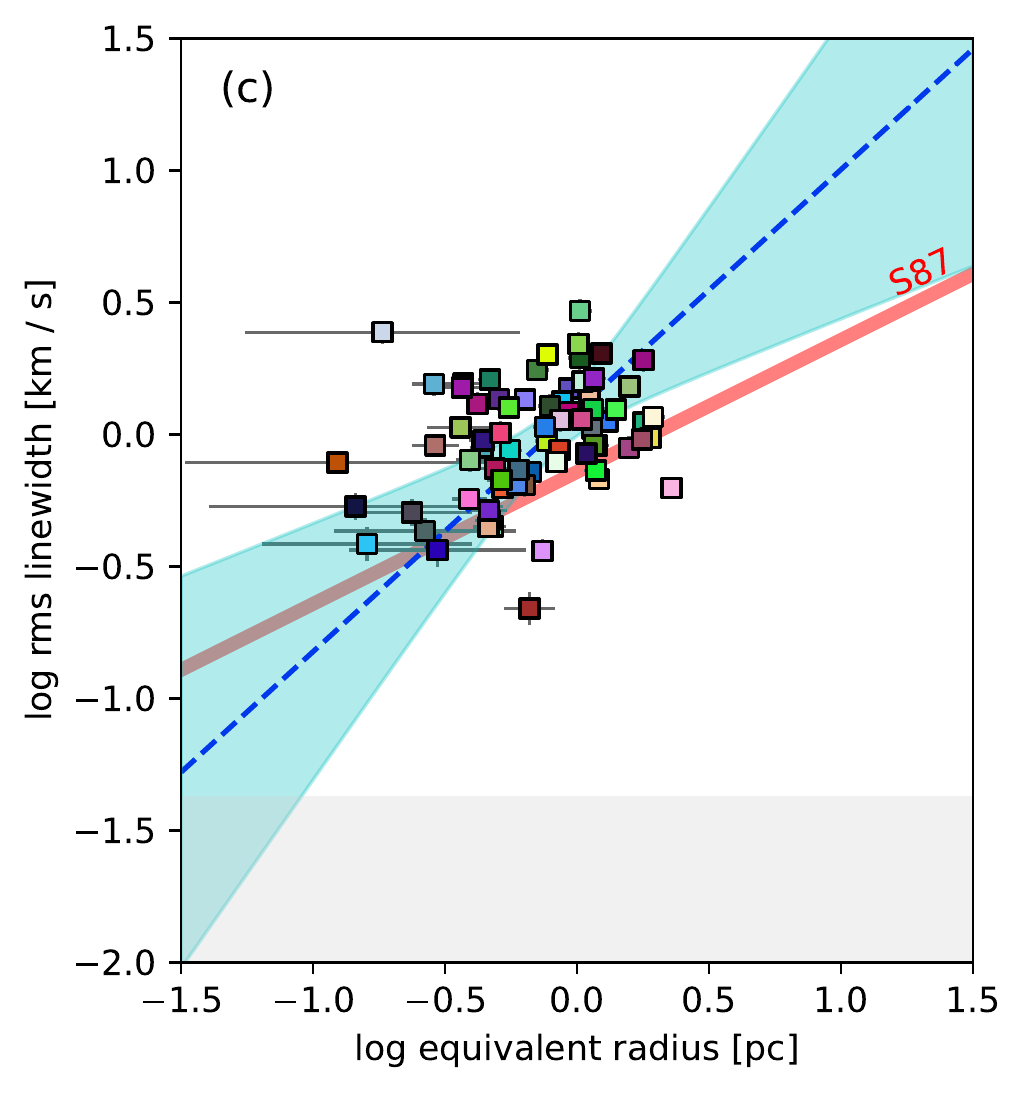}
\caption{Size-linewidth relations for SCIMES clumps identified in the feathered data: (a) \twco\ clumps; (b) \ttco\ clumps; (c) \ttco\ clumps at 0.1 \kms\ velocity resolution.  The power law fit and 3$\sigma$ uncertainty are shown in blue; {the gray shaded region indicates the limiting spectral resolution}. Fit parameters are tabulated in Tables~\ref{tab:fitpar} and \ref{tab:fitpar2}.
\label{fig:rdv_clusters}}
\end{figure*}

\section{Results}\label{sec:results}

\subsection{Overall cloud structure}\label{sec:overview}

Figures~\ref{fig:snrpk} and \ref{fig:mom01} show that the overall morphology of the cloud is primarily oriented along a direction rotated $\sim$30\arcdeg\ counterclockwise from north.  
{The left panel of Figure~\ref{fig:mom01} shows an overlay of the integrated CO intensity as magenta contours over a 3-color image (using the F555W, F775W, and F160W filters) from HTTP \citep{sabbi:13}, revealing that in some instances the CO is associated with extincted regions situated in the foreground of the Tarantula Nebula.}
As apparent from earlier single-dish mapping \citep{johansson:98,minamidani:08,pineda:09}, the brightest CO emission is distributed in two triangular lobes that fan out from the approximate position of R136, giving the cloud its characteristic ``bowtie-shaped'' appearance.  ALMA resolves these triangular lobes into radially oriented filaments (Figure~\ref{fig:clust}), providing another example of the ``hub-filament'' structure previously reported in the N159 \HII\ region that lies just south of 30 Dor \citep{fukui:19,tokuda:19}.  A third large CO-emitting region to the northwest, closer to Hodge 301, is also highly filamentary but with more randomly oriented filaments. 

In terms of velocity structure, the 30 Dor cloud spans a relatively large extent in velocity (approximately 40 \kms), compared to the typical velocity extent of $\sim$10 \kms\ seen in other LMC molecular clouds \citep{saigo:17,wong:19}.  Figure~\ref{fig:mom01} shows that the bowtie-shaped structure is primarily blueshifted with respect to the mean cloud velocity {($\bar{v} \approx 255$ \kms\ in the LSRK frame or $\bar{v}_\odot = 270$ \kms)}, with a relatively faint redshifted structure seen crossing perpendicular to it from the northwest to southeast.  The clouds projected closest to R136 and studied by \citet{kalari:18} are among the most highly blueshifted in the region and are observed in extinction against the \HII\ region, indicating that they are situated in the foreground.  The mean stellar velocity of the R136 cluster ($v_\odot = 271.6$ \kms; \citealt{evans:15}) is consistent with the mean cloud velocity, while the ionized gas has a somewhat lower mean velocity ($v_\odot = 267.4$ \kms; \citealt{torres:13}).

\subsection{Size-linewidth relations}\label{sec:rdv}

A correlation between size and line width, of the form $\sigma_v \propto R^{\gamma}$ with $\gamma \approx 0.5$, has long been observed among molecular clouds as well as their substructures \citep[hereafter \citetalias{solomon:87}]{larson:81,solomon:87}.  It is usually interpreted in the context of a supersonic turbulent cascade spanning a wide range of spatial scales \citep{maclow:04, falgarone:09}. The line width vs.\ size relations for the dendrogram structures in 30 Dor are summarized in Figures~\ref{fig:rdv_feather} and \ref{fig:rdv_clusters} for all structures and for the SCIMES clumps respectively.  {Gray shading indicates line widths which would be unresolved at the spectral resolution of the corresponding cube; nearly all of the significant structures are well-resolved in velocity.}
The standard relation of \citetalias{solomon:87} (with a slope and intercept of $a_1=0.5$ and $a_0=-0.14$ respectively) is shown as a thick red line for reference.  The best-fitting slopes and intercepts, derived using the {\tt kmpfit} module of the Python package {\tt Kapteyn}, are tabulated in Table~\ref{tab:fitpar}, {along with the reduced $\chi^2$ of the fit and the residual scatter along the $y$-axis}.  Consistent with previous studies (see \S\ref{sec:intro}), the relation in the 30 Dor cloud is offset to larger line widths compared to \citetalias{solomon:87}, by a factor of 1.5--1.8. The enhancement in line width we find is somewhat smaller than the factor of $\sim$2.3 previously derived for the ALMA Cycle 0 data \citep{nayak:16,wong:17}, indicating that the {central} region observed in Cycle 0 has a larger enhancement in line width than the cloud as a whole.  We revisit the positional dependence of the line width vs.\ size relation in \S\ref{sec:position}.

To evaluate the robustness of the fitted relations to the data handling procedures, we fit the relations separately for cubes derived from the 12m-only data and the feathered data, and for cubes with 0.1 \kms\ velocity channels and 0.25 \kms\ velocity channels.  
{The resulting fits are consistent within about twice the quoted 1$\sigma$ errors, as can be seen for example by comparing Tables~\ref{tab:fitpar} and Table~\ref{tab:fitpar2} and panels (b) and (c) of Figures~\ref{fig:rdv_feather} and \ref{fig:rdv_clusters}.  We note, however, that the fitted slope is often quite uncertain due to the limited range in structure size probed by our analysis, especially for the \ttco\ data.}

\begin{figure*}
\includegraphics[height=3.5in]{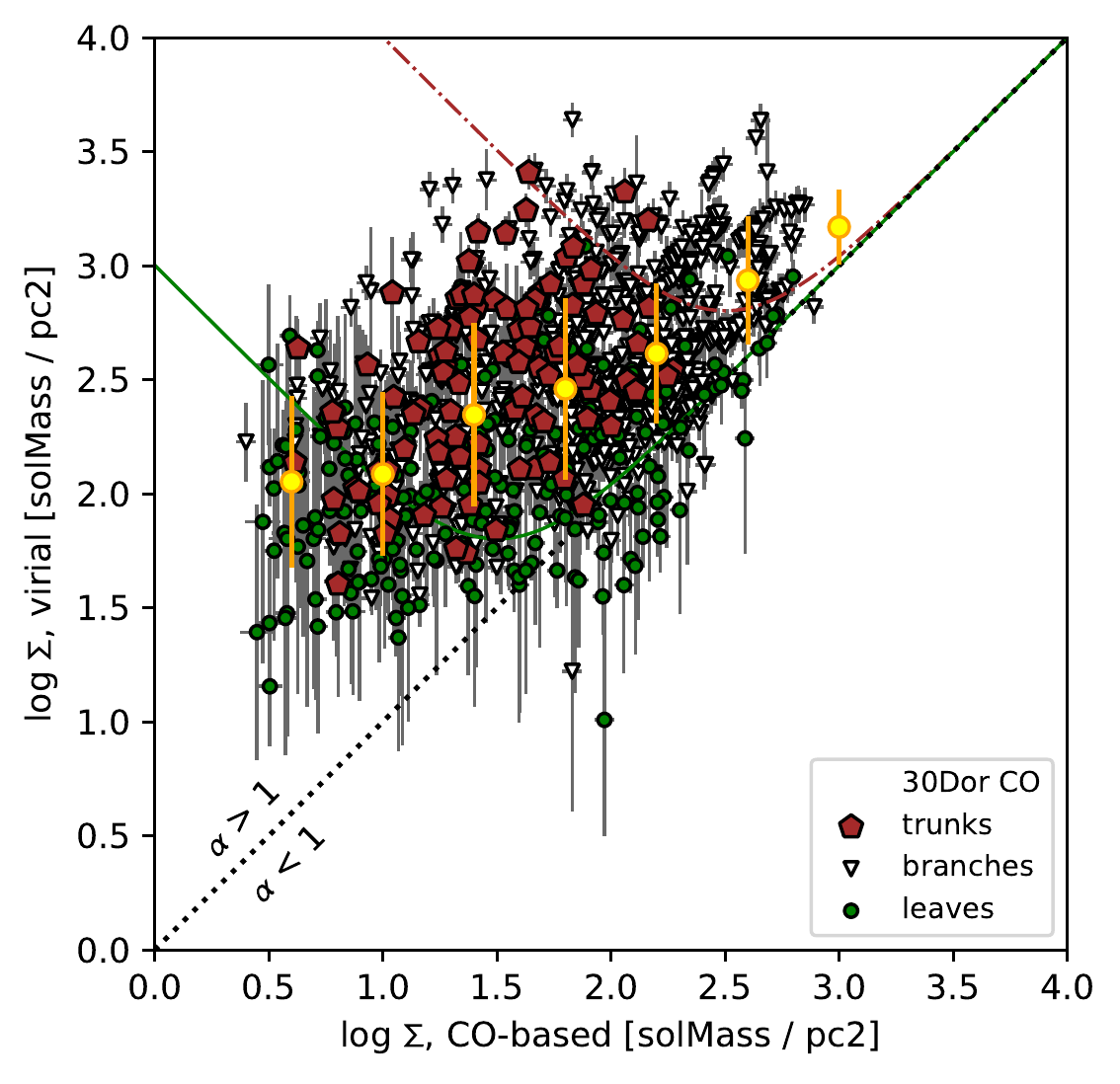}
\includegraphics[height=3.5in]{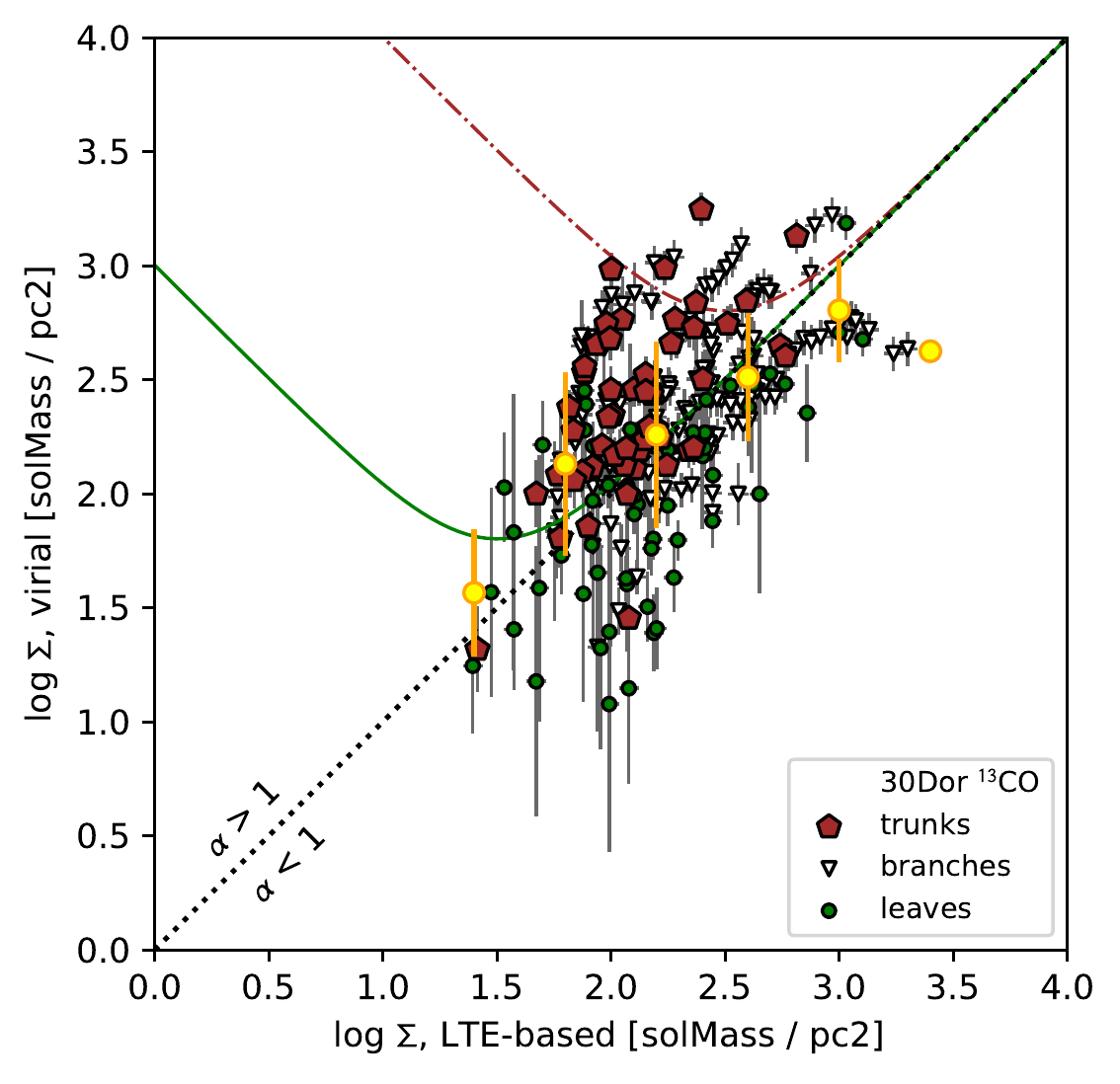}
\caption{Boundedness diagram for dendrogram structures identified in the feathered data.  {\it Left}: \twco\ structures, with surface density based on a constant $X_{\rm CO}$ factor. {\it Right}: \ttco\ structures, with surface density based on the LTE approximation.  Plot symbols indicate the type of dendrogram structure (trunks, branches, or leaves), with binned averages shown in yellow. The diagonal 1:1 line represents simple virial equilibrium, while the falling and rising solid green (dot-dashed red) curve represents pressure-bounded equilibrium with an external pressure of $10^4$ ($10^6$) cm$^{-3}$ K.}
\label{fig:bnd}
\end{figure*}

\begin{figure*}
\includegraphics[height=3.5in]{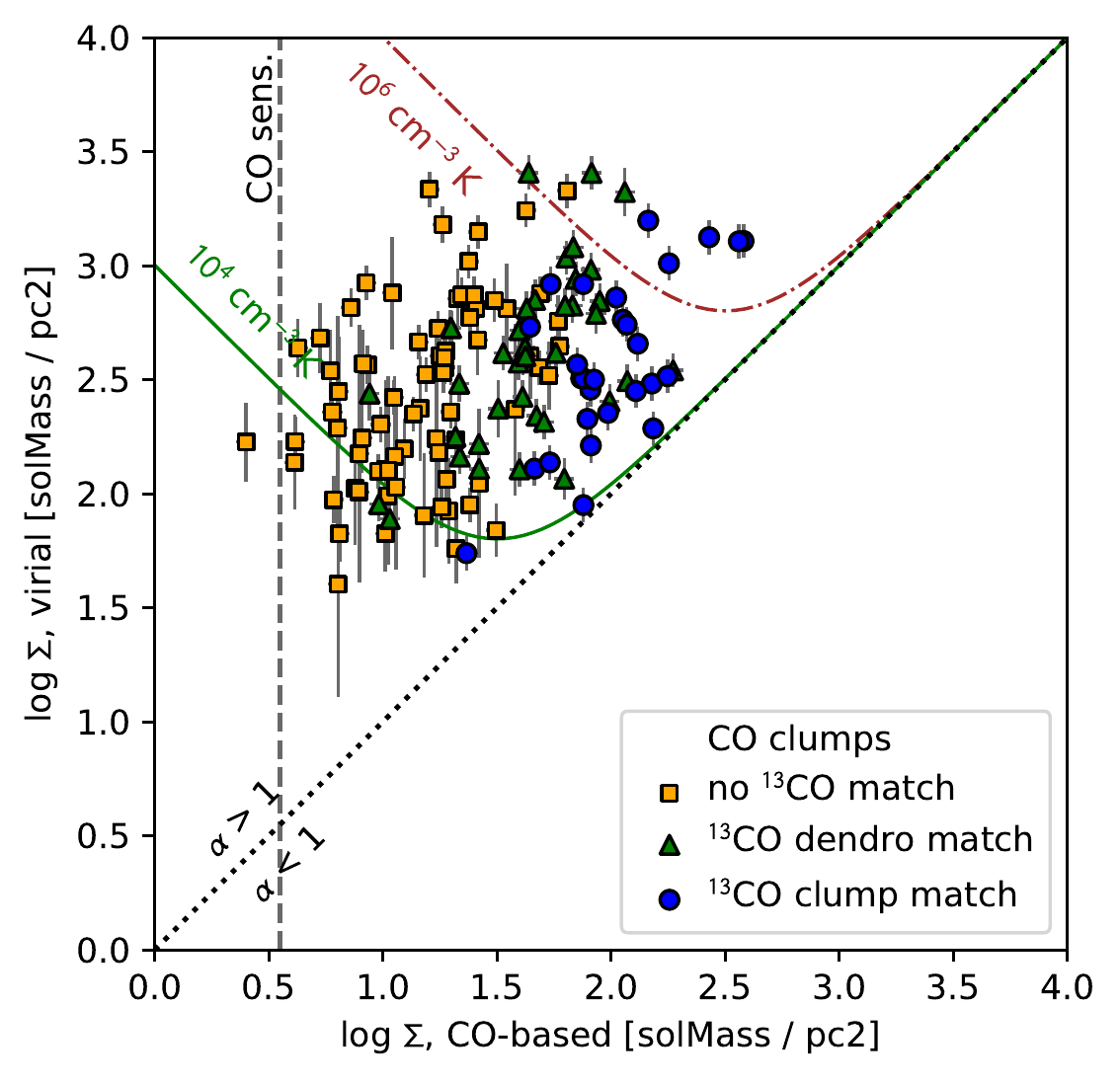}
\includegraphics[height=3.5in]{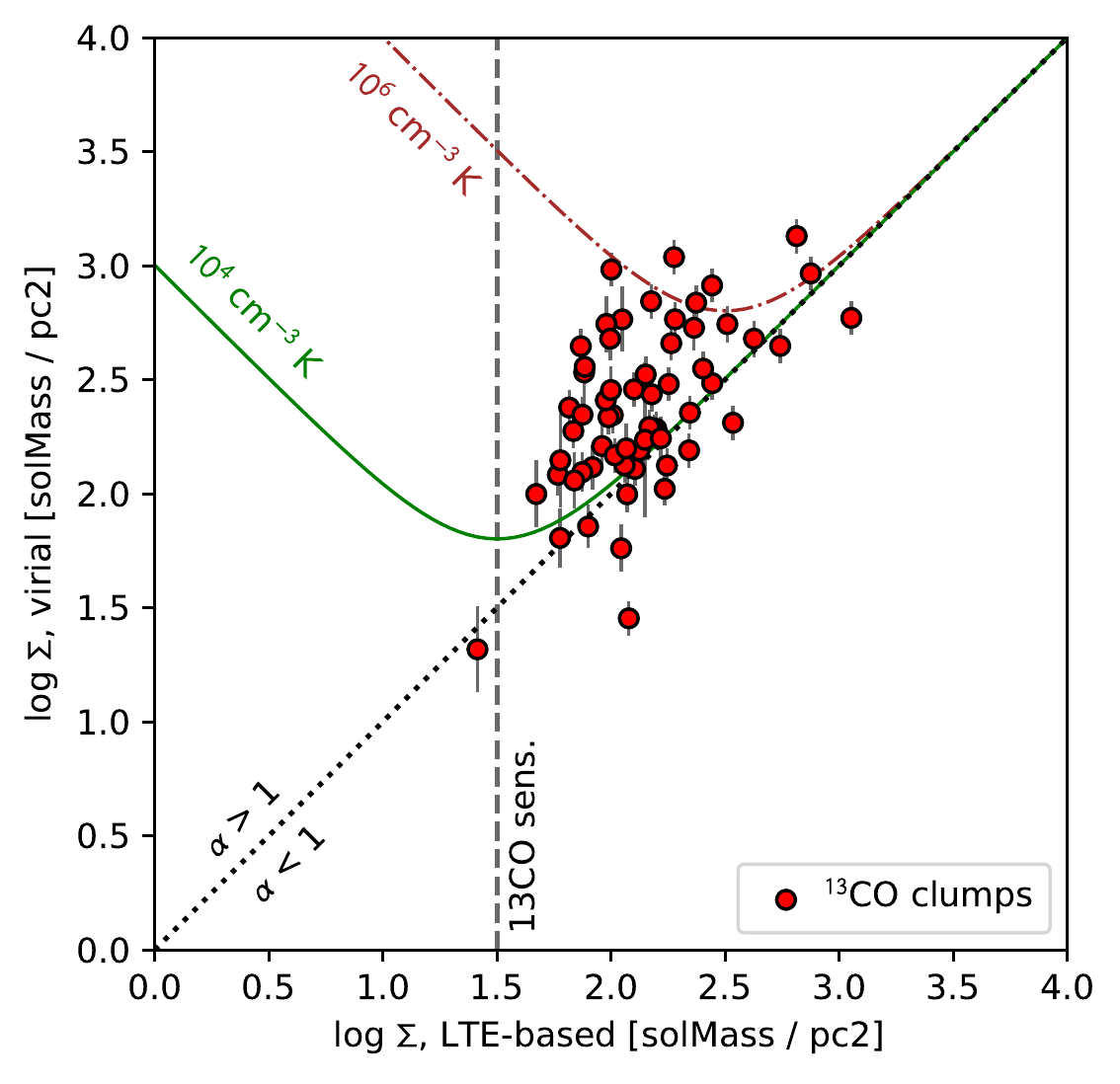}
\caption{Boundedness diagrams for SCIMES clumps identified in the feathered data.  Virial and pressure-bounded equilibrium curves are the same as in Figure~\ref{fig:bnd}. {\it Left}: \twco\ clumps, with surface density based on a constant $X_{\rm CO}$ factor.  Points are distinguished according to spatial overlap with any \ttco\ dendrogram structure (triangles) or \ttco\ clumps (circles). {\it Right}: \ttco\ clumps, with surface density based on the LTE approximation.  Vertical lines denote approximate 4$\sigma$ sensitivity limits for a 1 \kms\ line width; the \ttco\ sensitivity assumes $T_{\rm ex}$=8 K.}
\label{fig:bnd_clust}
\end{figure*}

\begin{figure*}
\includegraphics[height=3.3in,clip,trim=0 0 60 0]{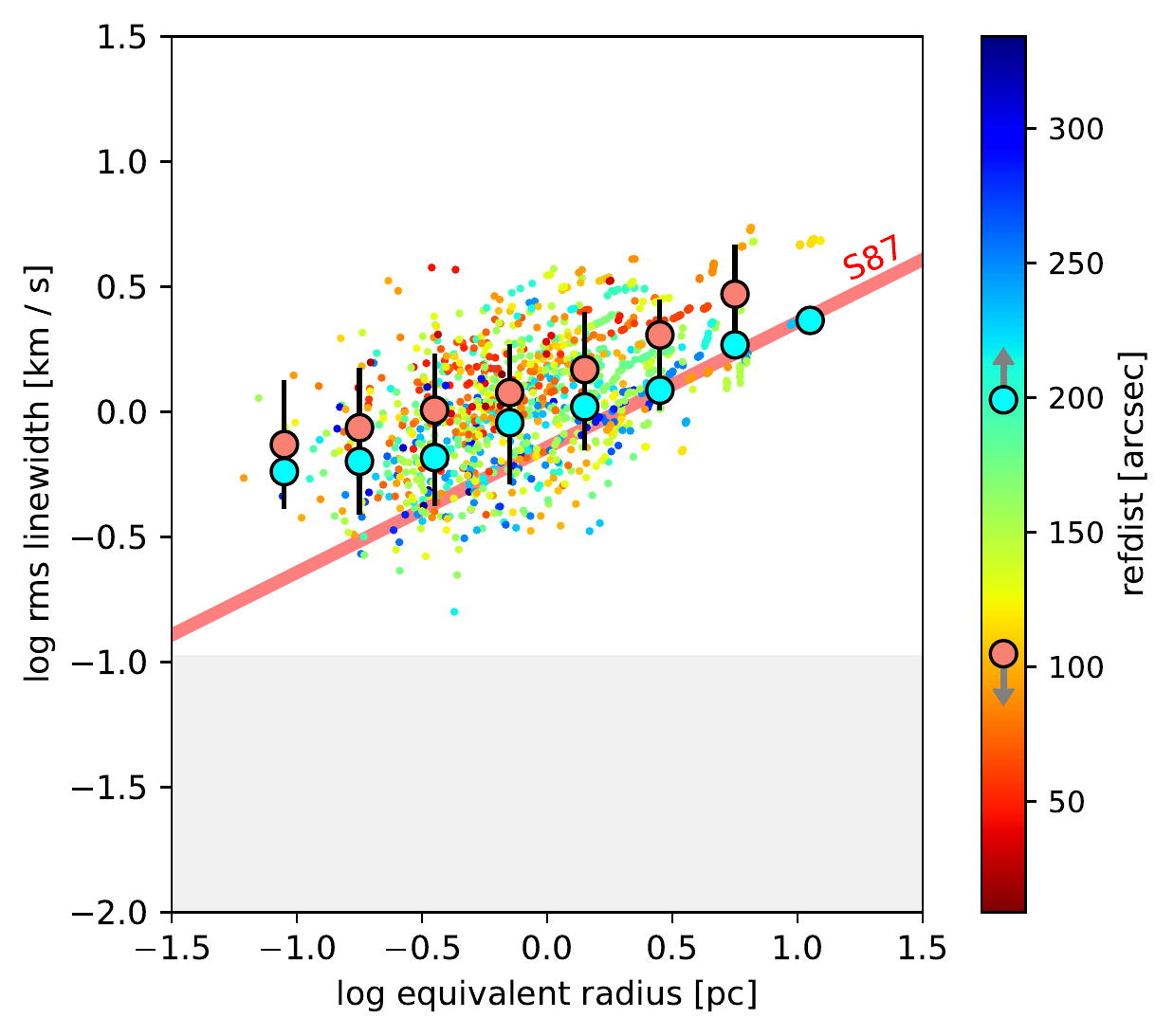}
\hfill
\includegraphics[height=3.3in]{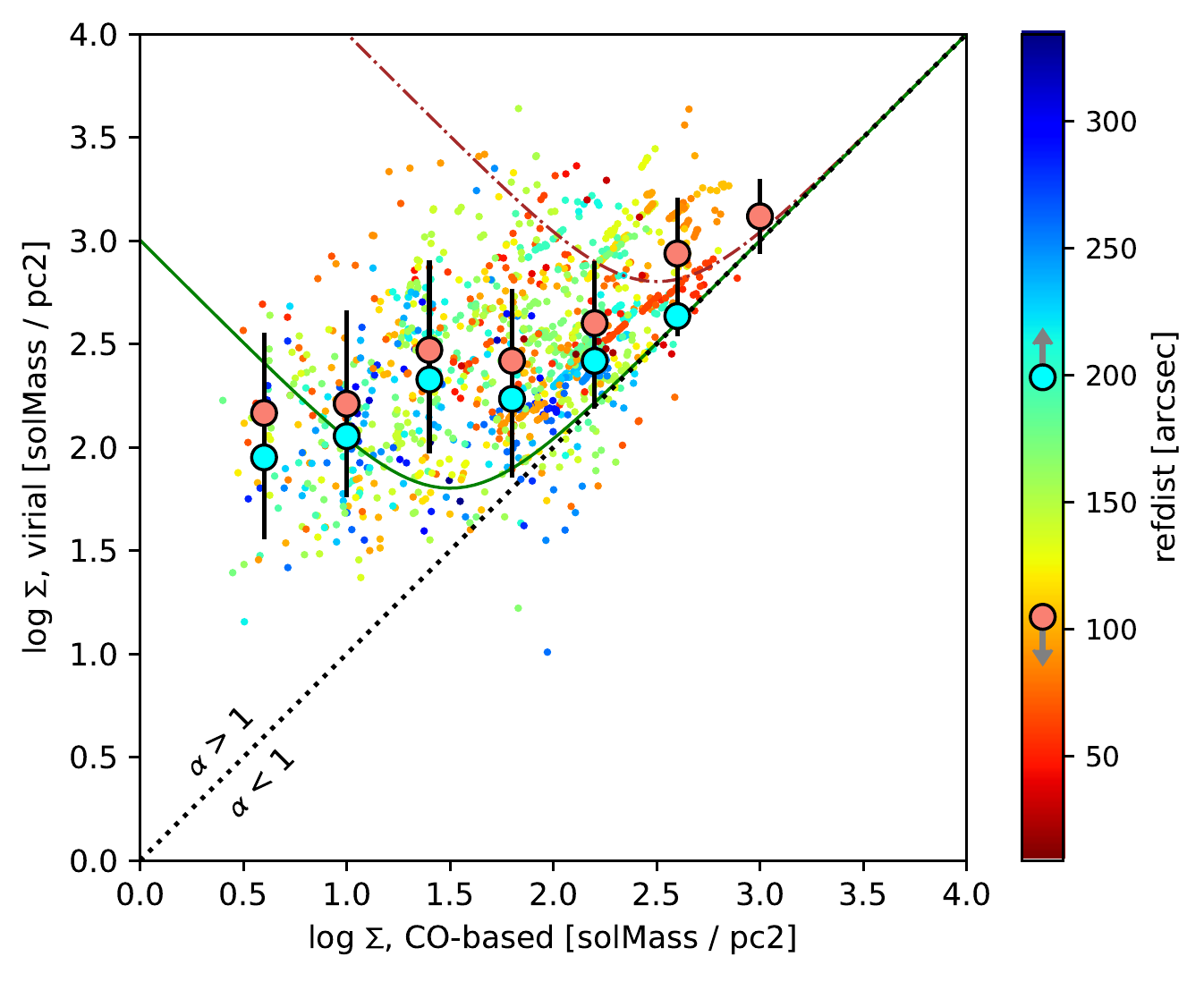}
\caption{Correlations between size and linewidth ({\it left}), and $\Sigma_{\rm vir}$ and $\Sigma_{\rm lum}$ ({\it right}), for the same \twco\ dendrogram structures plotted in Figures \ref{fig:rdv_feather} and \ref{fig:bnd}.  Distance from R136 is indicated by point colors and binned values (bins shown are averages of the top and bottom quartiles).  Since $\Sigma_{\rm vir} \propto \sigma_v^2/R$, higher line width at a given size results in higher $\Sigma_{\rm vir}$ for structures closer to R136.}
\label{fig:refdist12}
\end{figure*}

\begin{figure*}
\includegraphics[height=3.3in,clip,trim=0 0 60 0]{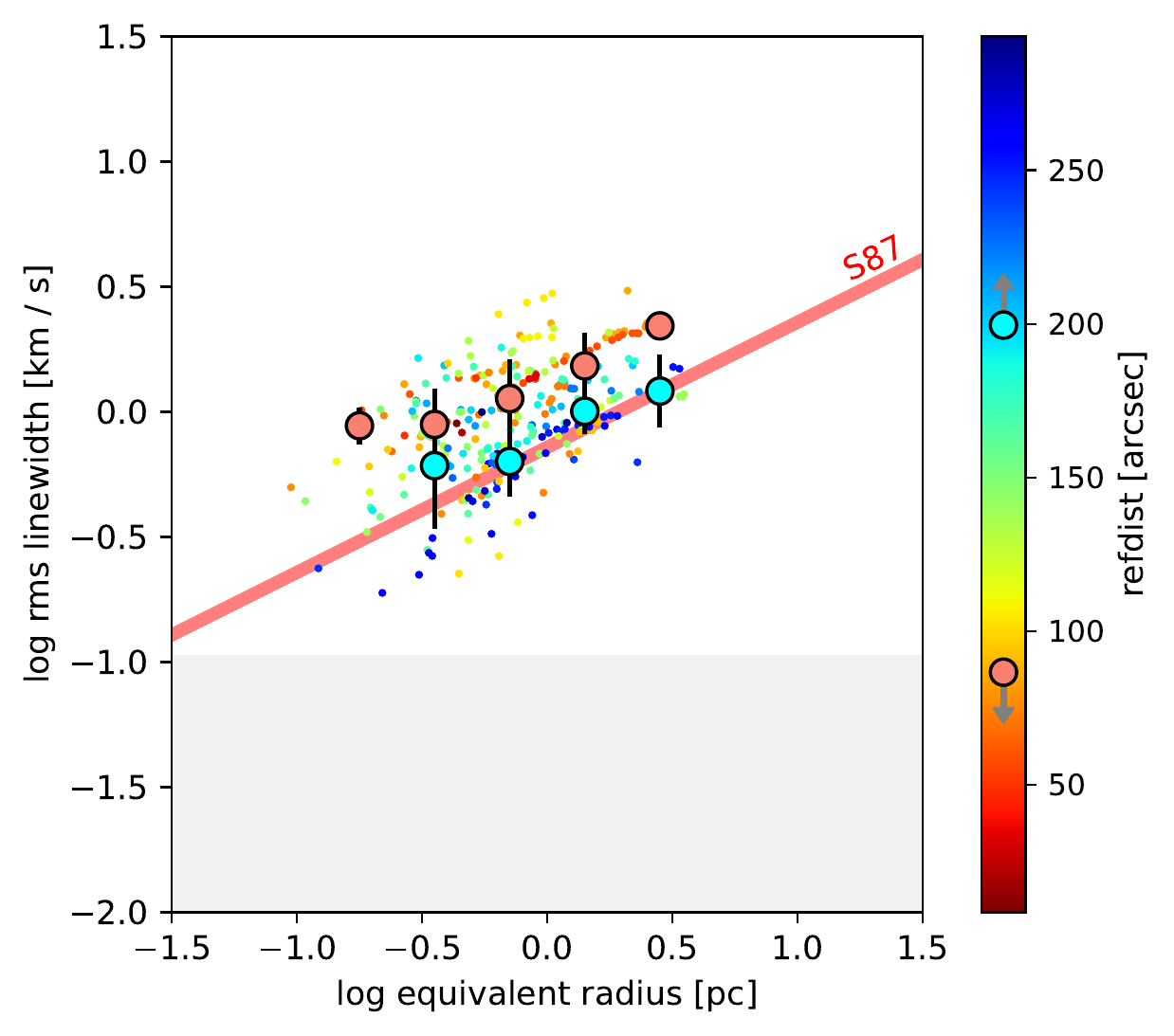}
\hfill
\includegraphics[height=3.3in]{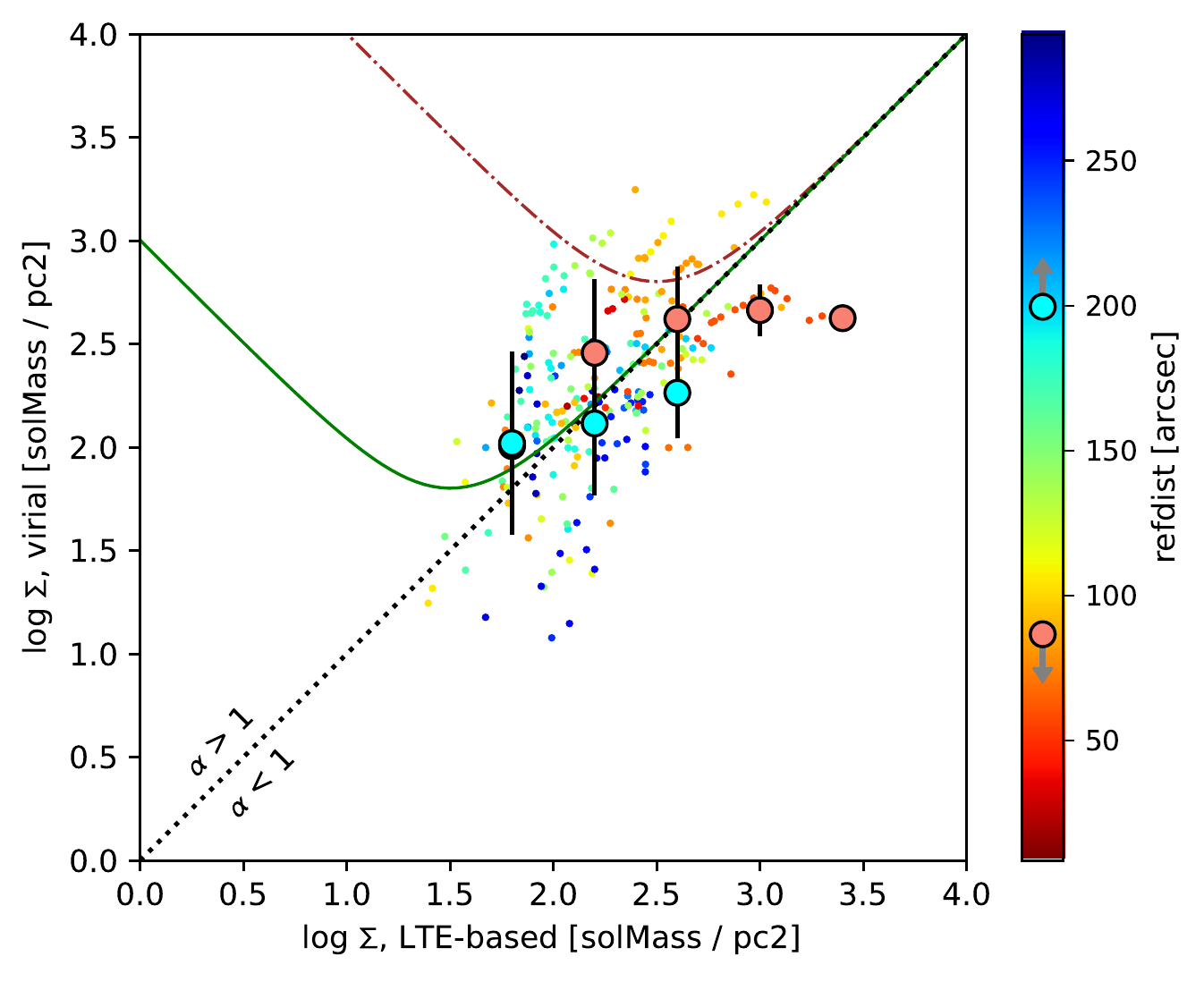}
\caption{Same as Figure~\ref{fig:refdist12}, but for \ttco\ dendrogram structures and with mass surface density based on the LTE approximation.}
\label{fig:refdist13}
\end{figure*}

\subsection{Virial relations}\label{sec:virial}

If the line width vs.\ size relation has a power-law slope of $\approx$0.5, then variations in the normalization coefficient $k$ are expected if structures lie close to virial equilibrium but span a range in mass surface density \citep{heyer:09}:
\begin{equation}
    \sigma_v = kR^{1/2} = 
    \left(\frac{\pi G}{5}\right)^{1/2}\Sigma_{\rm vir}^{1/2} R^{1/2} \quad\Rightarrow\; k = \sqrt{\frac{\pi G \Sigma_{\rm vir}}{5}}\;.
\label{eqn:heyer}
\end{equation}
This motivates an examination of whether variations in the line width vs.\ size coefficient are consistent with virial equilibrium.
For each structure whose {deconvolved} size and linewidth are {measured}, we normalize the virial and luminous mass by the projected area of the structure (determined by the pixel count) to calculate a mass surface density $\Sigma$.  For the \ttco\ structures, we use the LTE-based mass in preference to a \ttco\ luminosity-based mass, though the results tend to be similar.  The virial surface density, $\Sigma_{\rm vir}$, is directly related to the normalization of the size-linewidth relation, since $\Sigma_{\rm vir} = 5k^2/(\pi G)$ from Equation~\ref{eqn:heyer}.  We show the relations between $\Sigma_{\rm vir}$ and the luminous or LTE surface density in Figure~\ref{fig:bnd}.  In these ``boundedness'' plots, the $y=x$ line represents simple virial equilibrium (SVE), with points above the line having excess kinetic energy (often interpreted as requiring confinement by external pressure to be stable) and points below the line having excess gravitational energy (often interpreted as requiring support from magnetic fields to be stable).  

Overall, we find that \ttco\ structures are close to a state of SVE, with higher surface density structures tending to be more bound ($\avir = \Sigma_{\rm vir}/\Sigma_{\rm lum} \lesssim 1$).  On the other hand, \twco\ structures exhibit a shallower relation, with lower $\Sigma_{\rm lum}$ structures found to lie systematically above the SVE line.  The ``unbound'' CO structures exist across the dendrogram hierarchy ({spanning} leaves, branches, and trunks) and are found to dominate even the population of (typically larger) SCIMES clumps, as shown in Figure~\ref{fig:bnd_clust} (left panel).  
The mean value of $\log\avir$ for clumps without \ttco\ counterparts, as determined by checking for direct spatial overlap, is {1.26}, compared to {0.80} for clumps with \ttco\ counterparts (thus, the clumps detected in both lines have a factor of {3 lower} $\avir$).  

To better understand why the \twco\ structures appear less likely than \ttco\ structures to be bound, we need to bear in mind the sensitivity limitations imposed by the data. 
Most ({53\%}) CO clumps do not appear associated with \ttco, whereas all \ttco\ clumps overlap with a \twco\ clump.  
This reflects the fact that structures with lower CO surface brightness are less likely to be detected in \ttco: $\left<\log \Sigma_{\rm lum}\right> = 1.8$ for structures with \ttco\ counterparts while $\left<\log \Sigma_{\rm lum}\right>$ = {1.2} for those without \ttco\ counterparts.  A typical clump with a 1 \kms\ line width requires an integrated intensity of 0.55 K \kms\ to be detected at the 4$\sigma$ level.  As indicated by vertical dashed lines in Figure~\ref{fig:bnd_clust}, this intensity limit translates to minimum $\log \Sigma_{\rm lum} = 0.55$ for detection in \twco\ but a minimum $\log \Sigma_{\rm LTE} = 1.5$ for detection in \ttco\ (for $T_{\rm ex} = 8$ K).  Thus, the majority of \twco\ structures would not be expected to have \ttco\ counterparts because the weaker \ttco\ line was observed to the same brightness sensitivity as the stronger \twco\ line.  If lower surface density structures are preferentially unbound, then such structures will also tend to be detected only in \twco.

We note that several caveats apply to the interpretation of the ``boundedness'' plots.  As other authors have pointed out \citep[e.g.,][]{dib:07,ballesteros:11a}, 
objects that are far from equilibrium can still appear close to SVE as a result of approximate energy equipartition between kinetic and gravitational energies.  Furthermore, there are systematic uncertainties in estimating the values in both axes that are not included in the formal uncertainties.  For $\Sigma_{\rm vir}$ these include the spherical approximation and the definitions employed for measuring size and line width.  For $\Sigma_{\rm lum}$, uncertainties arising from the adoption of a single $X_{\rm CO}$ factor are ignored.  In particular, in regions with strong photodissociating flux it is possible for low column density \twco\ structures to be gravitationally bound by surrounding CO-dark gas (see \S\ref{sec:disc} for further discussion).  For $\Sigma_{\rm LTE}$, deviations from LTE conditions or errors in our assumed $T_{\rm ex}$ may affect the reliability of $\Sigma_{\rm LTE}$, although {from Equation~\ref{eqn:t13}} a shift in $T_{\rm ex}$ tends to be partially compensated by the resulting shift in $\tau_{13}$ and thus yield a similar value for $\Sigma_{\rm LTE}$.  An error in the assumed \ttco\ abundance would produce a more systematic shift, but would likely affect the cloud as a whole.

\begin{figure*}
\includegraphics[width=0.49\textwidth]{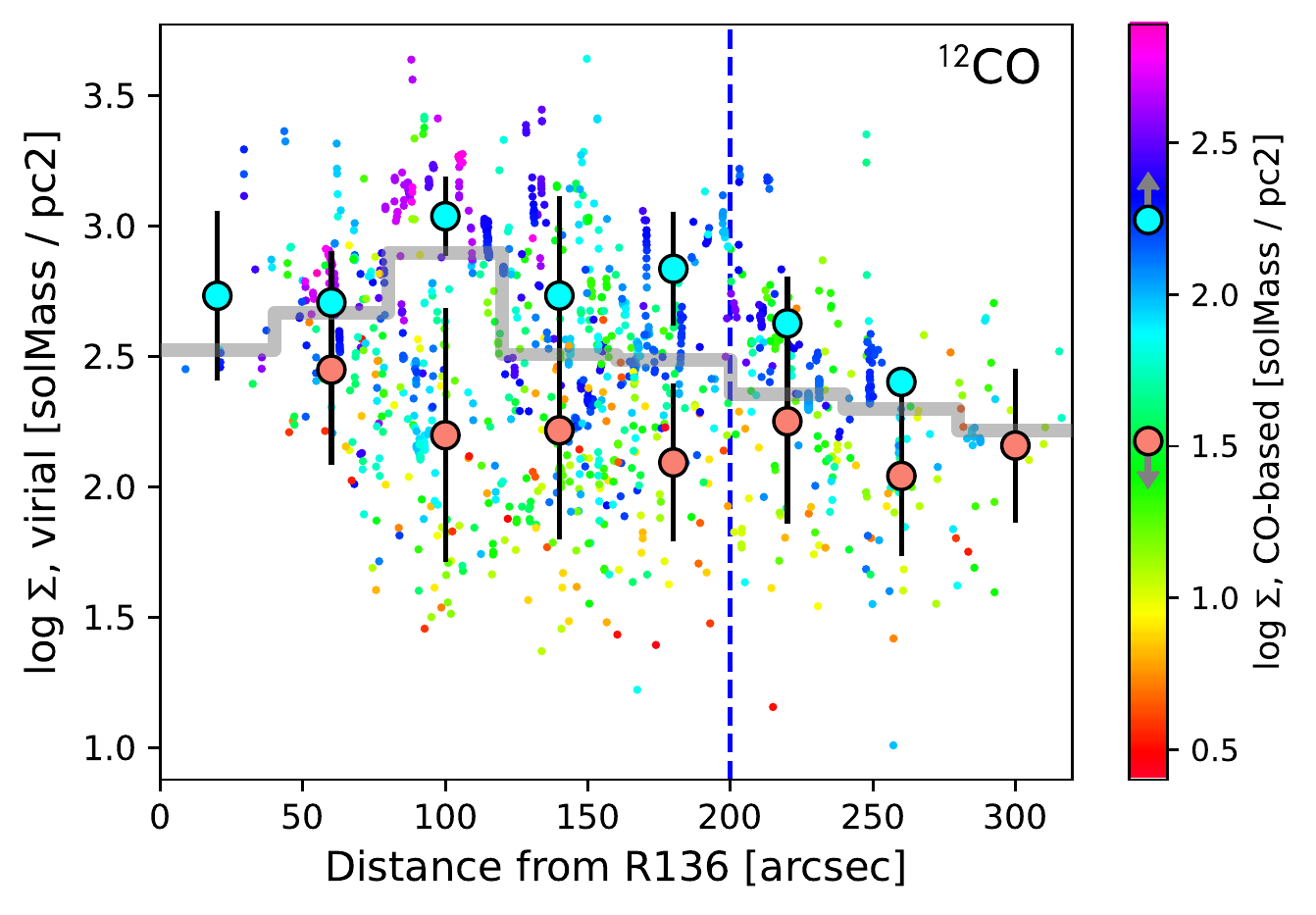}\hfill
\includegraphics[width=0.49\textwidth]{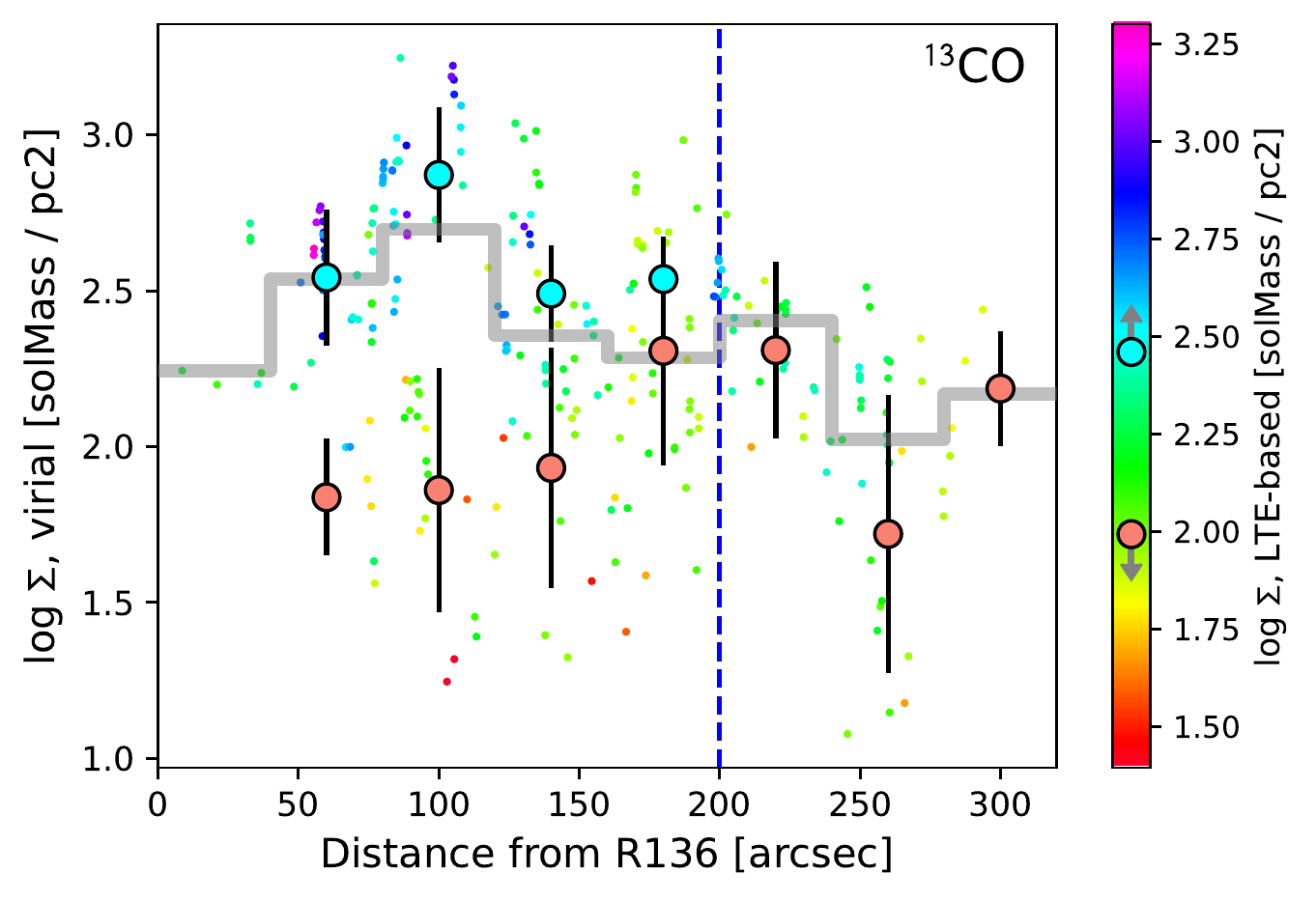}\\
\includegraphics[width=0.49\textwidth]{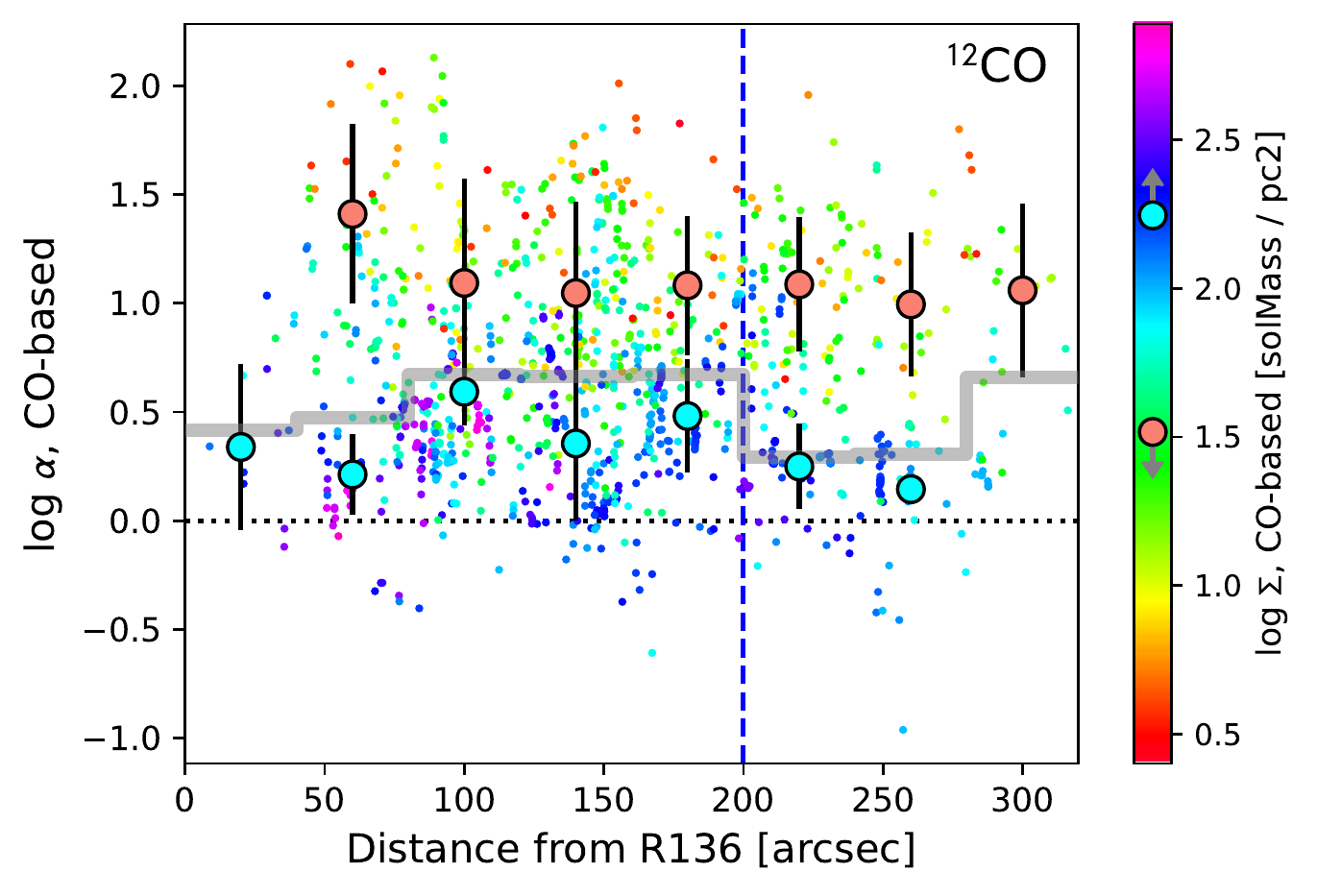}\hfill
\includegraphics[width=0.49\textwidth]{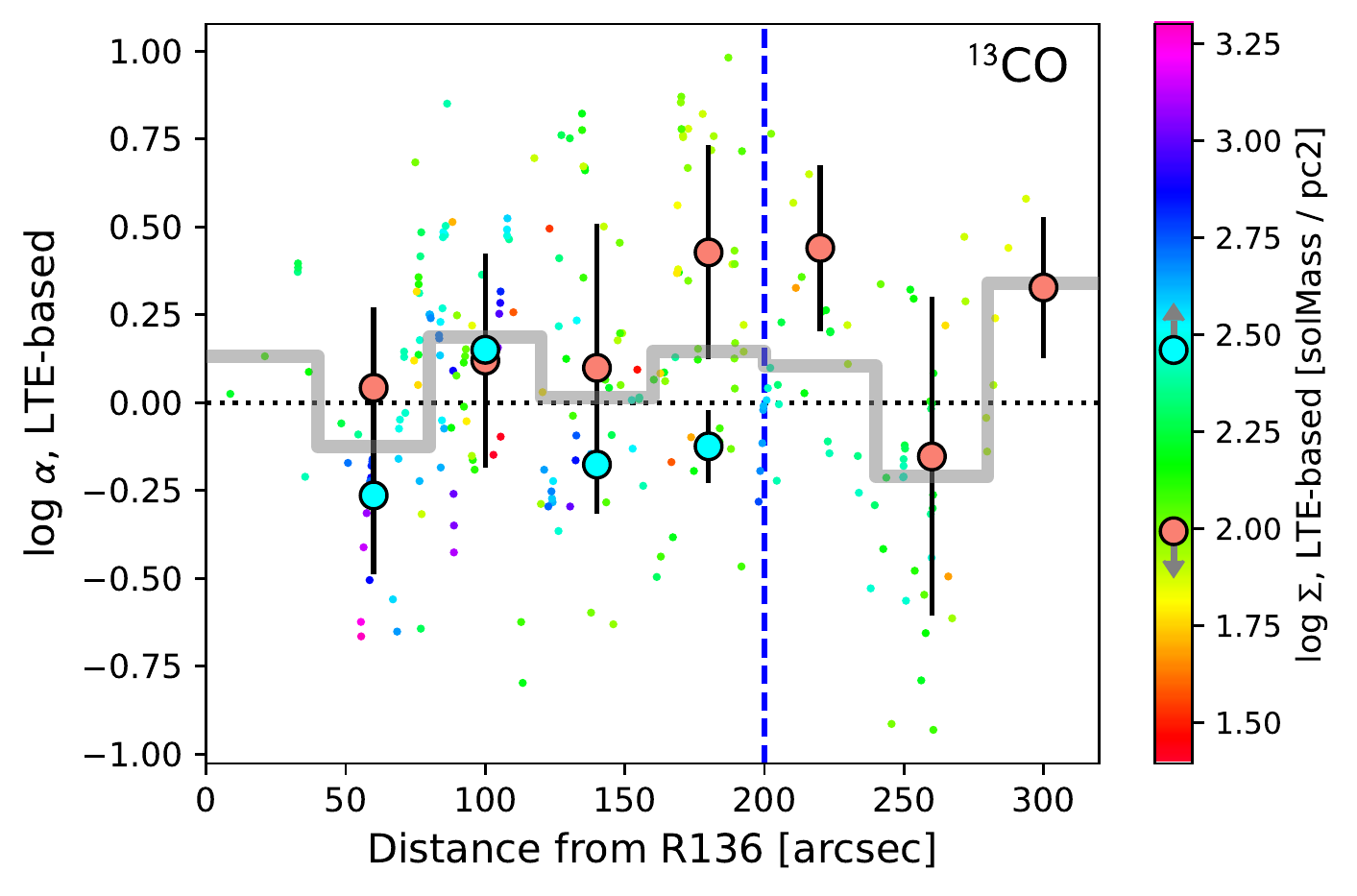}
\caption{Virial surface density $\Sigma_{\rm vir}$ ({\it top row}) and virial parameter $\alpha_{\rm vir}$ ({\it bottom row}) as a function of distance from R136 for \twco\ structures ({\it left}) and \ttco\ structures ({\it right}).  The colors of the plotted points represent mass surface density estimates, namely CO surface brightness for \twco\ and LTE column density for \ttco.  Binned values represent the highest and lowest 25\% of the {overall} mass surface density {and are plotted when two or more such points fall within a bin}.  Gray steps indicate the median value in each bin.  There is a decreasing trend in $\Sigma_{\rm vir}$ with distance, especially for the highest surface density structures, but no clear trend in $\alpha_{\rm vir}$.}
\label{fig:refdist_alpha}
\end{figure*}

\begin{figure*}
\includegraphics[width=\textwidth]{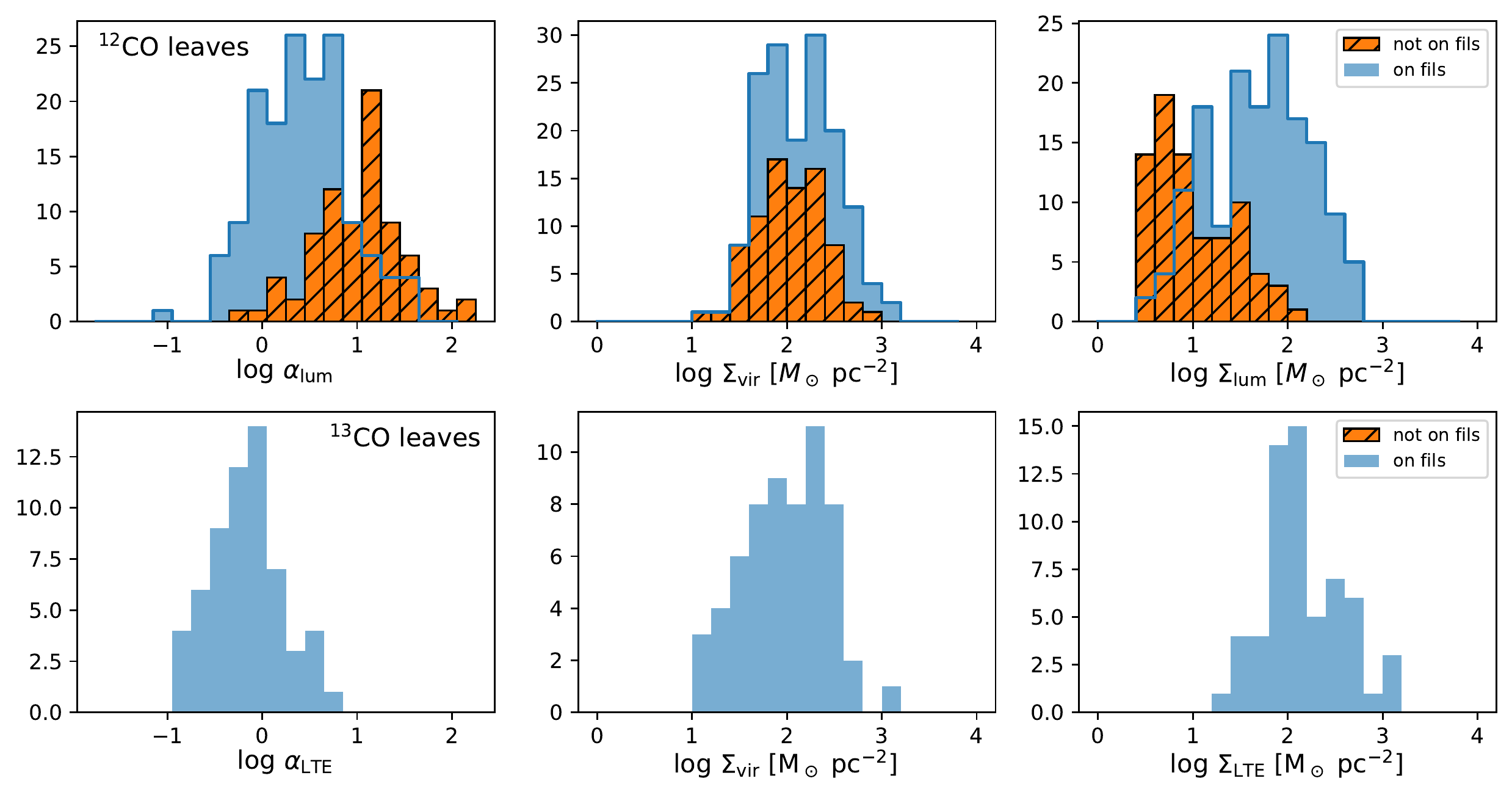}
\caption{Properties of leaf dendrogram structures {distinguished by positional coincidence with} \twco-identified filaments.  Note that histogram bars are superposed (rather than stacked) and unresolved structures have been excluded.  The top row shows the virial parameter $\avir$ and its constituent quantities $\Sigma_{\rm vir}$ and $\Sigma_{\rm lum}$ for the \twco\ leaves, whereas the bottom row shows the same for the \ttco\ leaves.  The \twco\ structures on filaments tend to have lower $\avir$ driven by higher surface density, whereas \ttco\ structures are {exclusively} found on filaments.}
\label{fig:leaf_fils}
\end{figure*}

\subsection{Position dependent properties}\label{sec:position}

To assess position-dependent variations in the size-linewidth and boundedness relations, we examine these relations color-coded by projected angular distance from the R136 cluster {($\theta_{\rm off}$ in Tables~\ref{tab:dendro12}--\ref{tab:clust13})} in Figures~\ref{fig:refdist12} and \ref{fig:refdist13}.  We also plot the binned correlations for the top and bottom quartiles of angular distance from R136.  We note that projected angular distance is only a crude indication of environment as it neglects the full 3-D structure of the region.  We find that regions at large angular distances are quite consistent with the \citet{solomon:87} size-linewidth relation {(except for the smallest structures, which have large uncertainties in the deconvolved size)}, whereas regions at smaller distances lie offset above it, consistent with previous studies \citep{indebetouw:13,nayak:16,wong:19}.  The approximate offset between the lowest and highest quartile of distances, at a fiducial size of 1 pc, is 0.16 dex (factor of 1.4) for \twco\ and {0.22} dex (factor of {1.7}) for \ttco.  As noted in \S\ref{sec:rdv}, an even larger (factor of $\sim$2) offset is found if one restricts the analysis to the Cycle 0 field.

When it comes to gravitational boundedness, the picture is more complex.  Structures close to R136 show higher $\Sigma_{\rm vir}$ in Figures~\ref{fig:refdist12} and \ref{fig:refdist13}, as expected given that $\Sigma_{\rm vir}$ scales with the size-linewidth coefficient $k$.  However, they exhibit no tendency to be more or less bound: \twco\ structures with low $\Sigma_{\rm lum}$ show excess kinetic energy relative to SVE at {\it all} distances from R136.  Figure~\ref{fig:refdist_alpha} provides a closer look at trends in $\Sigma_{\rm vir}$ and $\avir$ with distance from R136.  High surface density structures, represented by cyan circles, are close to virial equilibrium ($|\log\avir| \lesssim 0.5$) at all distances but tend to be concentrated towards R136, largely accounting for the higher $\Sigma_{\rm vir}$ observed in the central regions.  Beyond 200\arcsec\ from R136
{(to the right of the vertical dashed line), high surface density structures are largely absent.}
Meanwhile, the low surface density \twco\ structures, represented by red circles, are unbound ($\log\avir \gtrsim 1$) at all distances from R136.
{The median value of $\log\avir$ (represented by the gray steps) is largely unchanged with distance.}

\subsection{Association with filaments}\label{sec:filassoc}

{Galactic studies that have surveyed dense prestellar cores at far-infrared or submillimeter wavelengths \citep[e.g.,][]{fiorellino:21} have demonstrated a strong positional association of dense cores with filaments.  Here we conduct a preliminary assessment of this association in 30 Dor by comparing the dendrogram leaf structures to the filament skeleton derived by FilFinder.}
We present histograms of $\avir$, $\Sigma_{\rm vir}$, and $\Sigma_{\rm lum}$ (and their analogues in \ttco) for the leaf structures in Figure~\ref{fig:leaf_fils}, {distinguishing leaves by whether or not their actual structure boundaries (not their fitted Gaussians) overlap with the FilFinder skeleton.  Such overlaps must be viewed cautiously as both the structures and the filaments are identified using the same data set. Indeed, the SCIMES clumps are largely coincident with the FilFinder skeleton (Figure~\ref{fig:clust}).  In contrast, the \twco\ leaves constitute a large set of independent structures, and given their small typical sizes, a substantial fraction ($\sim$1/3) are not coincident with the skeleton, allowing us to compare the properties of leaves located on and off of filaments.}  
{Not surprisingly, the} filament-associated leaves tend to have higher $\Sigma_{\rm lum}$; in total they represent {93\%} of the total mass in leaves.  However, their values of $\Sigma_{\rm vir}$ are very similar to those of leaves which are not on filaments, and as a result the leaves on filaments tend to have lower $\avir$ (stronger gravitational binding).  The formation of filaments is therefore plausibly related to gravity, a hypothesis supported by the {fact that \ttco\ leaves}---which trace higher density material---{are exclusively associated} with the \twco\ filaments.

{Further analysis of the FilFinder outputs will be deferred to a future paper where we will collectively examine the properties and positional associations of YSOs, dense clumps, and filaments.}

\section{Discussion and Conclusions}\label{sec:disc}

We have presented initial results from an ALMA mosaic of CO(2--1) and \ttco(2--1) emission from the molecular cloud associated with the 30 Dor \HII\ region in the LMC, expanding upon the Cycle 0 map areal coverage by a factor of $\sim$40.  The emission exhibits a highly filamentary structure (Figures~\ref{fig:snrpk} and \ref{fig:clust}) with many of the longest filaments oriented radially with respect to ``hub'' regions nearer the cloud center.  The cloud's relatively large velocity width is resolved into several distinct components, with the bulk of the emission at lower radial velocity (Figures~\ref{fig:fluxcomp} and \ref{fig:mom01}).  We find that structures at a given size show decreasing line width with increasing distance from the central R136 cluster (Figures~\ref{fig:rdv_feather} and \ref{fig:rdv_clusters}), such that at the largest distances the normalization of the line width vs.\ size relation is consistent with the Galactic clouds studied by \citetalias{solomon:87}. However, we do not find that distance from R136 correlates with the gravitational boundedness of structures (Figure~\ref{fig:refdist_alpha}).  Rather, low surface density \twco\ structures tend to be unbound, whereas high surface density structures (which more closely follow the filamentary network, Figure~\ref{fig:leaf_fils}, and comprise most of the structures observed in \ttco) tend to be bound.  The higher line widths of clumps near R136 then largely reflect the higher surface density of clumps in this region.

While the unbound (high \avir) clumps are found throughout the cloud and are not limited to the smallest ``leaves'' in the dendrogram hierarchy, they tend not to overlap the filament skeletons, suggesting a more diffuse structure or distribution.
In total, 12\% of the total CO-based mass in SCIMES clumps is located in clumps with $\log\avir>1$, whereas {44\%} 
of the mass is in clumps with $\log\avir<0.5$.  Here we briefly discuss three possible interpretations of the high \avir\ structures.

\paragraph{Pressure-bounded structures}
In super star cluster-forming environments such as the Antennae galaxy merger \citep{johnson:15,finn:19}, massive molecular clouds are observed with virial masses well above the SVE line, implying large external pressures ($P/k_B \sim 10^8$--$10^9$ cm$^{-3}$ K) in order to be in equilibrium.  Although the estimated \HII\ region pressure of $\sim 10^{-9}$ dyn cm$^{-2}$ or $P/k_B \sim 7 \times 10^6$ cm$^{-3}$ K in the 30 Dor region \citep{lopez:11} would be sufficient to confine the observed $\avir>1$ clumps (Figure~\ref{fig:bnd_clust}), the distribution of points in the Figures~\ref{fig:bnd} and \ref{fig:bnd_clust} is not consistent with a constant external pressure, but rather suggests a smoothly increasing virial parameter with decreasing surface density. 
If instead there were large variations in external pressure, these would be expected to correlate with distance from R136 \citep{lopez:11}, but we do not find that the offset distance significantly affects boundedness (Figure~\ref{fig:refdist12}). 
We therefore view a pressure-bound equilibrium state to be a less likely scenario.

\paragraph{Dispersing molecular structures}
The unbound, low-column density \twco\ structures may represent molecular cloud material that exhibits excess kinetic energy as a result of being dispersed by energetic feedback.  The unusual concentration of massive stars in 30 Dor would then could account for the high frequency of such clumps, as similar column density ($1<\log\Sigma_{\rm lum}<2$) structures in other LMC clouds tend to lie closer to simple virial equilibrium \citep{wong:19}. 
A crude estimate of the total kinetic energy (${\cal T} = 3M_{\rm lum}\sigma_v^2$) in {\twco} clumps with $\log \alpha > 1$ is $7 \times 10^{48}$ erg.  Using the estimate of mechanical stellar wind feedback from R136 of $1.2 \times 10^{39}$ erg s$^{-1}$ from \citet{bestenlehner:20}, it would take only $\sim$200 yr for R136 to inject this amount of energy.  (For comparison, the total kinetic energy in all clumps is $7 \times 10^{49}$ erg, with a corresponding time scale of $\sim$2000 yr.)  This suggests that stellar feedback could easily account for the excess line widths seen in the unbound structures, even if the coupling of the feedback energy into the molecular cloud motions is relatively inefficient.
The energetic feedback should preferentially and effectively disrupt low column density structures, as few such structures lie near the SVE line.

\paragraph{Massive CO-dark envelopes}
If there is a substantial amount of hidden molecular mass which is not traced by \twco\ or \ttco\ emission; i.e.\ ``CO-dark'' gas, low CO intensities may disguise considerably larger column densities, and overall virial equilibrium may still hold once the additional mass is accounted for.  The basis of this scenario \citep[see][and references therein]{chevance:20} is efficient CO photodissociation relative to H$_2$, since the latter is able to self-shield whereas CO is mainly shielded by dust.  Since 30 Dor is both a metal poor {\it and} highly irradiated environment, the amount of CO-dark gas may be substantial, especially for clouds or clumps where the total gas column density is low.  This effect is clearly illustrated in \citet[][Figure 20]{jameson:18}, where at low $A_V$ the $X_{\rm CO}$ factor is increased by approximately an order of magnitude compared to the Galactic value.  In the 30 Dor region, based on PDR modeling of far-infrared emission lines, \citet{chevance:20} conclude that the $X_{\rm CO}$ factor is enhanced by factors of 4--20 compared to the Galactic value.  Correcting for this enhancement would increase $\log\,\Sigma_{\rm lum}$ by 0.4--1.1 (given our adopted $X_{\rm CO}$) and bring the low column density structures shown in Figures~\ref{fig:refdist_alpha} and \ref{fig:leaf_fils} closer to virial equilibrium.  We caution, however, that the virial surface density $\Sigma_{\rm vir}$ is also affected by the underestimate of $R$ {and $\sigma_v$} resulting from CO-dark gas; {the net effect on \avir\ depends sensitively on the adopted density and velocity dispersion profiles within the clumps \citep{oneill:22}}.  In addition, the CO-dark gas would need to be preferentially distributed in low column density clouds, since the high column density clouds do not show an excess of apparent kinetic energy.

Future studies are still needed to test these interpretations and to place 30 Dor in the context of its larger environment and the LMC as a whole.  Wider-field imaging with ALMA should be able to incorporate regions which are outside the reach of massive star feedback and examine the consequences for clump properties.
In addition, detailing the extent and contribution of the CO-dark gas (e.g., using [\CI] and [\CII] mapping) over a sample of molecular clouds with matched CO mapping will clarify the effects that this component may have on the observed properties of CO clumps.

Images and data products presented in this paper are available for download from the Illinois Data Bank at \url{doi.org/10.13012/B2IDB-1671495_V1}.

\begin{acknowledgments}
{We thank the anonymous referee for helpful suggestions that substantially improved the paper.}
This paper makes use of the following ALMA data: ADS/JAO.ALMA \#2019.1.00843.S. 
ALMA is a partnership of ESO (representing its member states), NSF (USA) and NINS (Japan), together with NRC (Canada), NSC and ASIAA (Taiwan), and KASI (Republic of Korea), in cooperation with the Republic of Chile.
The Joint ALMA Observatory is operated by ESO, AUI/NRAO and NAOJ.
The National Radio Astronomy Observatory is a facility of the National Science Foundation operated under cooperative agreement by Associated Universities, Inc.
T.W., M.M., and R.I. acknowledge support from collaborative NSF AAG awards 2009849, 2009544, and 2009624.
A.D.B. acknowledges support from NSF AAG award 2108140.
M.R. acknowledges support from ANID (Chile) FONDECYT grant No.\ 1190684 and partial support from ANID Basal projects ACE210002 and FB210003.
{K.T. acknowledges support from NAOJ ALMA Scientific Research grant Nos. 2022-22B, and Grants-in-Aid for Scientific Research (KAKENHI) of Japan Society for the Promotion of Science (JSPS; grant Nos., JP21H00049, and JP21K13962).}
This project has received funding from the European Research Council (ERC) under the European Union's Horizon 2020 research and innovation programme (Grant agreement No. 851435). The authors acknowledge assistance from Allegro, the European ALMA Regional Center node in the Netherlands.

\end{acknowledgments}

\facilities{ALMA, HST}

\software{CASA \citep{mcmullin:07}, {\tt astrodendro} (\url{http://www.dendrograms.org}), Kapteyn (\url{https://kapteyn.readthedocs.io}), FilFinder \citep{koch:15}, SCIMES \citep{colombo:15}, Astropy \citep{astropy:13, astropy:18}, APLpy \citep{aplpy:12}.}

\bibliographystyle{aasjournal}
\bibliography{alma30dor.bib}

\end{document}